\documentclass[prb,aps,amssymb,nofootinbib,twocolumn,showpacs]{revtex4}
\usepackage{amsmath}
\usepackage{amssymb}
\usepackage{amsthm}
\usepackage{amsfonts}
\usepackage{enumerate}
\usepackage{latexsym}
\usepackage[dvips]{graphicx}

\newcommand{\beq}{\begin{equation}}
\newcommand{\eneq}{\end{equation}}
\newcommand{\beqs}{\begin{equation*}}
\newcommand{\eneqs}{\end{equation*}}

\input{epsf}

\begin{document}

\tolerance 1000

\title{Topological Excitations and their Contribution to Quantum
Criticality in $2+1$ D Antiferromagnets}

\author { Zaira Nazario$^\dagger$ and
David I. Santiago$^{\dagger, \star}$ }

\affiliation{$\dagger$ Department of Physics, Stanford University,
             Stanford, California 94305 \\
             $\star$ Gravity Probe B Relativity Mission, Stanford,
             California 94305}
\begin{abstract}
\begin{center}

\parbox{14cm}{ It has been proposed that there are new degrees of
freedom intrinsic to quantum critical points that contribute to
quantum critical physics. We study $2+1$ D antiferromagnets in order
to explore possible new quantum critical physics arising from
nontrivial topological effects. We show that skyrmion excitations are
stable at criticality and have nonzero probability at arbitrarily low
temperatures. To include quantum critical skyrmion effects, we find a
class of exact solutions composed of skyrmion and antiskyrmion
superpositions, which we call topolons. We include the topolons in the
partition function and renormalize by integrating out small size
topolons and short wavelength spin waves. We obtain a correlation
length critical exponent $\nu=0.9297$ and anomalous dimension
$\eta=0.3381$.}

\end{center}
\end{abstract}
\date{\today}

\maketitle

\footnotetext{Z.N. is now at the Max Planck Institute for the Physics of
Complex Systems in Dresden, Germany. D.I.S. is now at the
Instituut-Lorentz at Leiden University in Leiden, The Netherlands.}

There have been recently interesting suggestions that there are
intrinsic degrees of freedom at quantum critical points and that these
intrinsic critical excitations are different from the elementary
excitations the quantum critical point separates\cite{bob1,
sachdev2}. In order to explore a path toward this possible new physics
at quantum critical points, we study the approach to the quantum
critical point from the N\'eel ordered phase of $2+1$ D
antiferromagnets. A physics that is usually not included in the
studies of criticality is the effects of the nontrivial topology of
the $O(3)$ group on the quantum critical physics. In the present
article we study this topology and include its effects on the critical
physics. We also calculate its contribution to critical exponents. In
the end we assess how close or how far we are from finding the
effective field theory of the critical point deriving from its
intrinsic critical excitations.

\section{$2 + 1$ D O(3) Nonlinear Sigma Model in the 
Stereographic Projection Language}

We now turn to $2+1$ D antiferromagnets whose effective field theory
is given by the nonlinear $\sigma$ model augmented by Berry phase
terms\cite{hal2,subir}:
    \begin{align}
      \begin{aligned}
        \mathcal Z &= \int \mathcal D\vec n \delta(\vec n^2-1) e^{-S}
	\\
        S = S_B + \frac{\rho_s}{2} &\int_{0}^{\beta} \!  d\tau \!
	\int \!  d^2\vec x \left[(\partial_{\vec x}\vec n)^2 +
	  \frac{1}{c^2}(\partial_\tau\vec n)^2 \right] \\
	= S_B + \frac{\Lambda}{2 g} &\int_{0}^{\beta} \! d(c \tau)
	\!  \int \!  d^2\vec x \left[(\partial_{\vec x}\vec n)^2 +
	  \frac{1}{c^2}(\partial_\tau\vec n)^2 \right]
      \end{aligned}
    \end{align}
where $a$ is the lattice constant, $\Lambda = 1/a$, $S$ is the
microscopic spin with $\hbar$ included (not to be confused with the
Euclidean action), and $J$ is the microscopic spin
exchange. $\rho_s\equiv J S^2/\hbar$ is the spin stiffness, $c = 2
\sqrt{2} J S a$ is the spin-wave velocity, and the dimensionless
microscopic coupling constant is $g = 2 \sqrt{2} / S$. These values
are obtained for the case of nearest neighbor Heisenberg interactions
only. For this case, in order to move the system from a N\'eel ordered
phase ($g << 1$) to a disordered or quantum paramagnetic phase ($g >>
1$), we tune the microscopic spin from large to small values. In real
life one does not have this option as the microscopic spin is
fixed. Moreover, the smallest available spin $S = 1/2$ is not enough
to place the system in the quantum paramagnetic phase\cite{halp}. In
order to quantum disorder the system in real life, one would need to
add frustrating next nearest neighbor interactions $J'$, frustrating
ring exchange interactions $K$, or other longer range but still short
range interactions that compete with the nearest neighbor N\'eel order
interaction. In this case, the dimensionless coupling $g$ becomes a
function of the ratios of the different interactions $g = g(J/J', J/K,
\dots)$ and by tunning the competing interactions, one can take the
system from the N\'eel ordered to the quantum paramagnetic phase.

The Berry phase term is the sum of the areas swept by the vectors
$\vec n_i(\tau)$ on the surface of a unit sphere at each lattice site
as they evolve in Euclidean time\cite{hal2}. The Berry phase terms are
zero when the N\'eel magnetization is continuous, i.e. N\'eel ordered
phase and critical point, as the contributions from neighboring
lattice sites cancel\cite{hal2, nohopf, vector}. Since we will
concentrate on critical properties as approached from the N\'eel
ordered side, the Berry phase vanishes.

A particular parametrization of the nonlinear sigma model that will
prove specially convenient when we shortly move to study the
topological structure of the model is obtained by mapping the
staggered magnetization $\vec n$ to the complex variable $w$ via the
stereographic mapping\cite{gross1}
    \beq
      n^1 \!+ i n^2 \!= \!\frac{2w}{|w|^2 \!+ \!1} \;, \; n^3 \!=
      \!\frac{1 \!- \!|w|^2}{1 \!+ \!|w|^2} \;, \; w \!= \!\frac{n^1
      \!+ \!i n^2}{1 \!+ \!n^3} \;.
    \eneq
More details of the stereographic projection are given in Appendix
\ref{topology}. In terms of $w$ the nonlinear sigma model action is
    \begin{align}
      \label{stereoact}
      S &= \frac{2 \Lambda}{g_\Lambda} \int d^3x \frac{\partial^\mu w
      \partial_\mu w^*}{ (1 + |w|^2)^2} \\
      \nonumber
      &= \frac{2 \Lambda}{g_\Lambda}\int d^3x \frac{\partial_0 w
      \partial_0 w^* + 2 \partial_z w \partial_{z^*} w^* + 2
      \partial_{z^*} w \partial_z w^*}{(1 + |w|^2)^2}
    \end{align}
where $\partial_0=\partial/\partial (c\tau)$. We have renamed $g$ as
$g_\Lambda$ because this is the microscopic coupling constant. In
order to study the phases of the model and the quantum critical point,
we will perform renormalization group studies\cite{wilson1, polya2,
br1, br2, wilson2, halp}. When we integrate degrees of freedom, the
coupling constants will become renormalized, but retain the same form
since the nonlinear sigma model is a renormalizable field
theory\cite{br1, br2, br3}. Thus the renormalized theory takes the
form
    \beq 
      S = \frac{2 \mu}{g_\mu} \int d^3x \frac{\partial^\alpha w
      \partial_\alpha w^*}{ (1 + |w|^2)^2}
    \eneq
with $\mu$ is the sliding renormalization scale\cite{weinberg} and
$g_\mu$ the renormalized coupling constant. We have suppressed field
renormalization factors.

\subsection{Traditional Goldstone Renormalization in the 
Stereographic Projection Language}

In order to gain experience with stereographic projections, to show
that the results are equivalent to more used approaches\cite{polya2,
br1, br2, wilson2, halp, sachdev3} and for later use we now perform a
one loop renormalization expansion of the nonlinear sigma model.

In terms of $w$, the partition function is
    \beq
      \mathcal Z = \int\prod_{\tau,\vec x}\frac{\mathcal D w(\tau,\vec
      x)\mathcal D w(\tau,\vec x)^*}{(1+|w(\tau,\vec x)|^2)^2}
      e^{-S}
    \eneq
where the Euclidean action $S$ is given by (\ref{stereoact}).  A
tricky point in the partition function is the nontrivial measure
arising from the nonlinearity of the sigma model and from the Jacobian
of the transformation between the $\vec n$ variables and the
stereographic projection $w$:
    \beq
      \prod_{\tau,\vec x}\frac{1}{(1+|w|^2)^2} =
      \exp{\left[-2\sum_{\tau,\vec x}\ln(1+|w|^2)\right]}
    \eneq
These products, or the sum in the exponentials, are not well defined
in the continuum limit. How this term is treated depends on your
regularization method. For example, in dimensional regularization this
term is just ignored with the price of some infrared divergences that
must be treated carefully. On the other hand, direct passage to the
continuum limit yields
    \begin{align}
      \begin{aligned}
	&\exp \left[-2\sum_{\tau,\vec x} \ln \left( 1 + |w|^2 \right)
	\right] = \\
	&\exp \left[ -2\delta^3(0) \int d^3 x \ln \left( 1 + |w|^2
	\right) \right] \;.
      \end{aligned}
    \end{align}
The delta functions are ill defined as they are infinite. This can be
dealt with in at least two equivalent ways, either by introducing a
lattice cutoff in real space or a momentum cutoff in momentum
space. We thus have
    \begin{align}
      \begin{aligned}
	&\exp{\left[-2\sum_{\tau,\vec x}\ln(1+|w|^2)\right]} = \\
	&\exp{\left[ -\frac{2}{a^3} \int d^3 x \ln(1+|w|^2)\right]}
      \end{aligned}
    \end{align}
where space integrals are cutoff at the short distance $a$. Then the
partition function can be written as
    \beq
      \mathcal Z = \int \prod_{\tau,\vec x} \mathcal D w(\tau,\vec x)
      \, \mathcal D w^*(\tau,\vec x) \, e^{-S_{eff}}
    \eneq
where the effective action is the nonlinear sigma model action
augmented by the terms obtained from the nontrivial measure:
    \beq
      S_{eff} = \frac{2}{a^3} \int d^3 x \ln \left( 1 + |w|^2 \right)
      + S \;.
    \eneq

We are interested in the N\'eel ordered phase and in the critical
point, but this last will be studied as it is approached from the
N\'eel ordered phase. So $g_\Lambda$ will not be too large and, at
least in the N\'eel ordered phase, a perturbative expansion in
$g_\Lambda$ is possible. Renormalization group resummations of this
expansion yield approximations to the critical theory. The lowest
order terms in $g_\Lambda$ will be given by the approximation
    \begin{align}
      \begin{aligned}
	S_{eff} &\simeq \frac{2}{a^3} \int d^3 x \, |w|^2 +
	\frac{2\Lambda}{g_\Lambda} \int d^3 x \, \partial^\mu w \,
	\partial_\mu w^* \\
	&- \frac{4\Lambda}{g_\Lambda} \int d^3 x \, |w|^2 \,
	\partial^\mu w \, \partial_\mu w^* \;.
      \end{aligned}
    \end{align}
We now work in momentum space. The effective action is then
    \begin{widetext}
      \begin{align}
	\begin{aligned}
          S_{eff} &= \frac{2 V^2}{a^3} \int \frac{d^3 k}{(2\pi)^3}
          |w(k)|^2 + \frac{2 V^2 \Lambda}{g_\Lambda} \int
          \frac{d^3k}{(2\pi)^3} \, k^2 |w(k)|^2 \\
	  &- \frac{4 V^4 \Lambda}{g_\Lambda} \int \frac{d^3 k_1 d^3
          k_2 d^3 k_3}{(2\pi)^9} \, w(k_1) w^*(k_2) \, k_3 \cdot (k_1
          - k_2 + k_3) \, w(k_3) w^*(k_1 - k_2 + k_3)
	\end{aligned}
      \end{align}
    \end{widetext}
where $k = (\omega, \vec k)$ and $V$ is the volume where the system
lies. We will slightly change our definition of the cutoff in order to
absorb some angular factors. Hence we renormalize the momentum sphere
to
    \beq
      V_3^\Lambda = \frac{4 \pi}{3} \Lambda^3 = \frac{1}{a^3} \;.
    \eneq

We'll renormalize the theory via momentum shell integration. In order
for this action to be well defined, i.e. finite, it is cutoff at large
momentum $\Lambda$ and we'll integrate the degrees of freedom from
$\Lambda$ to a smaller cutoff $\mu$. The bare action is
    \beq
      S_0 = 2 V^2 \frac{\Lambda}{g_\Lambda}\int \frac{d^3 k}{(2\pi)^3}
      \; k^2 |w(k)|^2
    \eneq
which leads to the bare momentum space Green's function or propagator
    \beq
      G_0 = \frac{(2\pi)^3}{2 V^2} \frac{g_\Lambda}{\Lambda}
      \; \frac{1}{k^2} \;.
    \eneq
The interaction term is given by
    \begin{align}
      \begin{aligned}
	&S_I = 2 \, V^2 \, V_3^\Lambda \, \int \frac{d^3k}{(2\pi)^3}
	\, |w(k)|^2 \\
	&- \, 4 \, V^4 \, \frac{\Lambda}{g_\Lambda} \int \left\{
	\frac{d^3k_1 \, d^3k_2 \, d^3k_3}{(2\pi)^9} w(k_1) \, w^*(k_2)
	\right. \\
	&\left. \times k_3 \cdot (k_1-k_2+k_3) \, w(k_3) \,
	w^*(k_1-k_2+k_3) \right\} \;.
      \end{aligned}
    \end{align}
We know integrate the high momentum degrees of freedom to obtain the
effective theory at scale $\mu$. If the momentum shell is thin enough,
a perturbative expansion is justified and to lowest order in the
interaction we have
    \beq
      \mathcal Z \simeq \int_\Lambda \mathcal D w \, \mathcal D w^*
      \left( 1-S_I\right) e^{-S_0} \,.
    \eneq
After integrating the high energy degrees of freedom ($\mu<k<\Lambda$)
we obtain
    \beq
      \mathcal Z \simeq \int_\mu \mathcal D w \mathcal D w^*
      e^{-S_{eff}^\mu} \,.
    \eneq
where the effective action at scale $\mu$ is given by 
    \beq
      S_{eff}^\mu = S_0 + \langle S_I\rangle
    \eneq
and the average of $S_I$ is taken over the large momentum degrees of
freedom. The noninteracting action $S_0$ is the one obtained from
$S_0$ at scale $\Lambda$ by integrating out the degrees of freedom
with momenta between $\mu$ and $\Lambda$ and the irrelevant constant
term are thrown out as they only modify the overall normalization. Now
we evaluate the expectation value of $S_I$ term by term. We first
obtain
    \begin{align}
      \begin{aligned}
	\langle S_I^{(1)}\rangle &= 2 \, V^2 \, V_3^\Lambda \, \int
	\frac{d^3k}{(2\pi)^3} \, \langle |w(k)|^2\rangle \\
	&= 2 \, V^2 \, V_3^\Lambda \int_{|k|<\mu} \frac{d^3
	k}{(2\pi)^3} |w(k)|^2 + C
      \end{aligned}
    \end{align}
$C$ is a constant; and
    \begin{widetext}
      \begin{align}
	\begin{aligned}
	  \langle S_I^{(2)} \rangle &= - 4 \, V^4 \,
	  \frac{\Lambda}{g_\Lambda} \int \left\{ \frac{d^3k_1 d^3k_2
	  d^3k_3}{(2\pi)^2} \langle w(k_1) w^*(k_2) \, k_3 \cdot (k_1
	  - k_2 + k_3) \, w(k_3) \, w^*(k_1 - k_2 + k_3) \rangle
	  \right\} \\
	  &- 2 \, V^2 \left( V_3^\Lambda - V_3^\mu \right)
	  \int_{|k|<\mu} \frac{d^3k}{(2\pi)^3} |w(k)|^2 -
	  \frac{V^2}{\pi^2} \left[\Lambda - \mu \right] \int_{|k|<\mu}
	  \frac{d^3 k}{(2\pi)^3} \, k^2 \, |w(k)|^2 + C
	\end{aligned}
      \end{align}
where the constant terms will be neglected as they only provide a
change of overall normalization of the partition function. Putting
everything together we obtain
      \begin{align}
	\begin{aligned}
          S_{eff}^\mu &= 2 \, V^2 \, V_3^\mu \int_{|k|<\mu} \frac{d^3
	  k}{(2\pi)^3} \, |w(k)|^2 + 2 \, V^2
	  \frac{\Lambda}{g_\Lambda} \left\{ 1 - \frac{g_\Lambda}{2 \,
	  \pi^2} \left[ 1 - \frac{\mu}{\Lambda} \right] \right\}
	  \int_{|k|<\mu} \frac{d^3 k}{(2 \pi)^3} \, k^2 \, |w(k)|^2 \\
	  &- 4 \, V^4 \, \frac{\Lambda}{g_\Lambda} \int_{|k_i|<\mu}
	  \frac{d^3k_1 \, d^3k_2 \, d^3k_3}{(2\pi)^9} \, \left\{ \,
	  w(k_1) \, w^*(k_2) k_3 \cdot (k_1 - k_2 + k_3) \, w(k_3) \,
	  w^*(k_1 - k_2 + k_3) \, \right\} \\
	  &\simeq 2 \, V^2 \, V_3^\mu \int_{|k|<\mu}
	  \frac{d^3k}{(2\pi)^3} |w(k)|^2 + 2 \, V^2 \,
	  \frac{\mu}{g_\mu} \int_{|k|<\mu} \frac{d^3k}{(2\pi)^3} \,
	  k^2 \, |w(k)|^2 \\
	  &- 4 \, V^4 \, \frac{\mu}{g_\mu} \int_{|k_i|<\mu}
	  \frac{d^3k_1 \, d^3k_2 \, d^3k_3}{(2\pi)^9} \, \left\{ \,
	  w(k_1) \, w^*(k_2) k_3 \cdot (k_1 - k_2 + k_3) \, w(k_3) \,
	  w^*(k_1 - k_2 + k_3) \right\} \;.
	\end{aligned}
      \end{align}
    \end{widetext}
We thus see that, to lowest order, the action
renormalizes\cite{polya2, br1, br2, halp}, i.e. goes into itself when
fast degrees of freedom are integrated out. This result can be proved
to be true to all orders by means of the $O(3)$ Ward identity of the
$O(3)$ nonlinear sigma model\cite{br3}.

The renormalized coupling constant at scale $\mu$ is seen to be
    \beq
      \label{reng}
      g_\mu = \left( \frac{\mu}{\Lambda} \right) \frac{g_\Lambda}{1 -
      \frac{g_\Lambda}{2 \pi^2} \left[1 - \frac{\mu}{\Lambda} \right]}
      \;.
    \eneq

This expression exemplifies quite a bit of the physics of the
renormalization group and the nonlinear sigma model. We first see that
the coupling constant gets renormalized to different values at
different scales. From this expression we see at least two fixed point
values of the coupling constant in the sense that for these two values
the coupling constant is the same at all momentum scales. One is
$g_\Lambda = g_{IR}^N = 0$ and the other is at $g_\Lambda = g_{UV}^c =
2 \pi^2$ as found long ago\cite{polya2}. $g_{IR}^N$ corresponds to the
N\'eel ordered or Goldstone phase and it is an infrared fixed point
which corresponds to a stable phase of matter. It is an infrared fixed
point because if $0< g_\Lambda < g_{UV}^c$, $g_\mu \rightarrow
g_{IR}^N =0$ as $\mu\rightarrow 0$. That is, as the energy scale is
lowered, the theory approaches the N\'eel ordered behavior.

The critical point where N\'eel order is lost corresponds to
$g_{UV}^c$. That this is a critical point follows since it is an
infrared unstable fixed point. If $g_\Lambda$ is arbitrarily close but
not exactly $g_{UV}^c$, it deviates from $g_{UV}^c$ as $\mu\rightarrow
0$. This is easily seen from our formula if $g_\Lambda$ is chosen to
be a value infinitesimally smaller than $g_{UV}^c$. More importantly,
the critical point is an ultraviolet fixed point because if
$0<g_\Lambda<g_{UV}^c$, $g_\mu\rightarrow g_{UV}^c$ as $\mu\rightarrow
\infty$. All critical points are ultraviolet fixed points. Therefore,
the critical properties can be studied from the N\'eel ordered phase
by studying their high momentum or high energy behavior. Finally, the
nonlinear sigma model has another infrared fixed point at $g_\Lambda =
\infty = g_{IR}^P$, which corresponds to the paramagnetic phase. This
fixed point cannot be accessed by expanding about the N\'eel ordered
phase as it is not adiabatically continuable to the N\'eel ordered
phase. In order to access this paramagnetic fixed point, one needs to
perform a strong coupling expansion. 

The spin stiffness of the nonlinear sigma model is proportional to the
inverse coupling constant
    \beq
      \rho_s \propto \frac{\mu}{g_\mu} \,.
    \eneq
Classically, $g_\mu/\mu = g_\Lambda/\Lambda$ is a constant and does
not get renormalized. The spin stiffness only vanishes when the bare
coupling constant $g_\Lambda$ becomes infinite. When fluctuation
effects are included, $g_\mu$ becomes renormalized according to
    \beq
      \rho_s \propto \frac{\Lambda}{g_\Lambda} \left[1 -
      \frac{g_\Lambda}{2 \pi^2} \left( 1 - \frac{\mu}{\Lambda} \right)
      \right]
    \eneq
If we tune $g_\Lambda$ to the critical value where N\'eel order is
lost, $g_{UV}^c$, the spin stiffness is
    \beq
      \rho_s \propto \frac{\mu}{g_{UV}^c} \,.
    \eneq
We thus see that at the quantum critical point, $g_\Lambda = g_{UV}^c
= 2\pi^2$, the spin stiffness vanishes at arbitrarily low energy
scales $\mu\rightarrow 0$.

The beta function is calculated from our expression for the
renormalized coupling constant $g$ as a function of $\mu$,
(\ref{reng}), to be

    \beq
      \beta(g) =\mu\frac{\partial g}{\partial\mu} \Big |_{\Lambda=\mu}
      = g - \frac{g^2}{2 \pi^2} \,.
    \eneq
The fixed points are given by the zeros of the $\beta$ function, which
are easily seen to occur at $g = 0$ and $g = 2 \pi^2$ as found
before. The first term, which is positive, arises because in more than
two space-time dimensions the theory is not truly scale invariant but
becomes so at fixed points. In more than two space-time dimensions,
the presence of the negative term makes possible the existence of a
critical point because at certain value of the coupling constant, the
negative or asymptotic freedom term cancels the positive scaling term
of the $\beta$ function.

Since our renormalization of the nonlinear sigma model was carried out
to one loop, we did not obtain the renormalization of the fields. The
situation is not as bad as it seems as it is possible to obtain a
field renormalization by a different one loop calculation. More
important than the renormalization of the $w$ fields, we are
interested in the renormalization $Z$ of the staggered
magnetization. Because of the $O(3)$ invariance, the Goldstone fields
$\Pi_1 = n_1$ and $\Pi_2 = n_2$, and the ordering direction or $\sigma
= n_3$ component are renormalized by the same factor $\sqrt{Z} =
\sqrt{1 + \delta} \simeq 1 + \delta/2$. We will fix the renormalized 
magnetization $\sigma$ to one in the N\'eel ordered phase. On the
other hand, the bare magnetization is related to the renormalized one
via $\sigma_0 = \sqrt{Z} \, \sigma \simeq (1 + \delta/2) \,
\sigma$. In order to obtain the renormalized magnetization we
calculate
    \begin{align}
      \begin{aligned}
	\sigma_0 &= \langle n_3\rangle = \left\langle
	\frac{1-|w(\tau,\vec x)|^2}{1+|w(\tau,\vec k)|^2}\right\rangle
	\simeq 1-2\langle |w(\tau,\vec x)|^2\rangle \\
	\sigma_0 &= \sqrt{Z} \simeq 1 + \frac{\delta}{2}
      \end{aligned}
    \end{align}
where the average value is obtained by integrating the fast degrees of
freedom between scales $\mu$ and $\Lambda$ in order to obtain the
magnetization at scale $\mu$. Therefore we obtain the counterterm
    \begin{align}
      \begin{aligned}
	\delta &= - 4 \langle |w(\tau,\vec x)|^2 \rangle = - 4 \, V^2
	\int_{\mu<|k|<\Lambda} \frac{d^3k}{(2\pi)^3} G_0(k) \\
	&= -\frac{g_\Lambda}{\pi^2} \left[ 1 - \frac{\mu}{\Lambda}
	  \right] \;.
      \end{aligned}
    \end{align}

$N$-point Green's functions of the nonlinear sigma model satisfy the
Callan-Symanzik (CS) equation\cite{osvia, weinberg, cs1, br1, br2}
    \beq
      \label{cs}
      \left[ \mu \frac{\partial}{\partial \mu} +
      \beta(g)\frac{\partial}{\partial g} + \frac{N}{2} \gamma(g)
      \right] G^{(N)}(p, g, \mu) = 0
    \eneq 
where the anomalous dimension $\gamma(g)$ is given by 
    \beq 
      \label{gamma}
      \gamma (g) = \mu \frac{\partial \ln Z}{\partial\mu}\Big
      |_{\mu=\Lambda} \simeq \mu\frac{\partial \delta}{\partial \mu}
      \Big |_{\mu=\Lambda} = \frac{g}{\pi^2}
    \eneq
We are particularly interested in the critical properties of
antiferromagnets and hence the nonlinear sigma model. When tuned to
criticality, we obtain that exactly at the UV fixed point or critical
point, $g_\Lambda = g_{UV}^c = 2 \pi^2$, the Green's function has the
form
    \beq 
      G^{(2)}(p, g_{UV}^c, \mu) = \frac{1}{p^2} \, h\left(
      \frac{p^2}{\mu^2}\right)
    \eneq 
which means that if we exchange the $\mu$ derivative for a $p$
derivative the CS equation becomes
    \beq
      \left[ p \frac{\partial}{\partial p} - \gamma(g_{UV}^c) \right]
      \, h\left(\frac{p^2}{\mu^2}\right) = 0 \;.
    \eneq 
This integrates immediately to 
    \beq 
      h \left( \frac{p^2}{h^2} \right) = A' \left( \frac{\mu^2}{p^2}
      \right)^{-\gamma/2}
    \eneq 
where $A'$ is an arbitrary integration constant. The two point
critical function is then
    \begin{align}
      \begin{aligned}
	G^{(2)}(p, g_{UV}^c, \mu) &= \frac{A'}{\mu^{\gamma(g_{UV}^c)}}
	\left( \frac{1}{p^2} \right)^{1 - \gamma(g_{UV}^c)/2} \\
	&\equiv \frac{A}{p^{2-\eta}} \,.
      \end{aligned}
    \end{align}

From equation (\ref{gamma}), we immediately find $\gamma(g_{UV}^c) =
2$ in agreement with previous one loop results\cite{polya2, br1}. This
is obviously wrong as the propagator loses all momentum
dependence. The reason for this nonsense is that higher order
corrections are quite large and must be included. More careful
approximations give less large and more sensical values\cite{br3,
halp, sachdev3}. We will evaluate below, for the first time, the
corrections coming from topological effects in order to get a more
accurate evaluation of the anomalous exponents.

Better than tuning the system exactly to criticality it is more
realistic to study the Green's function in the N\'eel ordered
phase. The two point Green's function in the N\'eel ordered phase
takes the form
    \begin{align}
      \begin{aligned}
	G^{(2)}(p, g, \mu) &= \frac{1}{p^2} h \left(
	\frac{\mu^2}{\Lambda^2} \right) = \frac{1}{p^2} h(g_\mu) \\
	&= \frac{1}{p^2} \, h \left( g, \frac{\mu^2}{p^2}\right) \,.
      \end{aligned}
    \end{align}
Because h is a function of $\mu^2/\Lambda^2$ only through the coupling
constant. Hence the CS equation can be integrated immediately to give
    \begin{align}
      \begin{aligned}
	G^{(2)} (p,g,\mu) &= \frac{A}{p^2} \exp{\left[
	-\int_{g_p}^{g_\mu} dg' \frac{\gamma (g')}{\beta (g')} \right]
	} \\
	&= \frac{A}{p^2} \left( 1 - \frac{g_\mu}{g_{UV}^c}\left[ 1 -
	\frac{p}{\mu} \right] \right)^2
      \end{aligned}
    \end{align}
We now see that the Green's function interpolates between the
Goldstones behavior at long wavelengths, $p\rightarrow 0$
    \beq
      G^{(2)} (p,g,\mu) = \frac{A}{p^2} \left( 1 -
      \frac{g_\mu}{g_{UV}^c} \right)^2
    \eneq
and critical behavior with anomalous exponent $\eta = 2$ at short
distances, $p\rightarrow \infty$.
    \beq
      G^{(2)} (p,g,\mu) = \frac{A}{p^2} \left[ \frac{g_\mu}{g_{UV}^c}
      \right]^2 \frac{p^2}{\mu^2} = \frac{A}{\mu^2} \, \left[
      \frac{g_\mu}{g_{UV}^c} \right]^2
    \eneq

Despite the large incorrect value of $\gamma$, the qualitative
conclusions obtained are true. There is a nonzero anomalous exponent
at the critical point and the Green's function satisfy the
Callan-Symanzik equation. Moreover, the conclusions drawn from the
Callan-Symanzik equation, i.e. the scaling behavior, becomes
quantitatively correct at small energy scales\cite{halp}. In fact, the
quantitative agreement between RG studies of the nonlinear sigma model
and neutron scattering results in the cuprate high T$_c$
superconductors demonstrated that these are N\'eel ordered at low
temperatures when sufficiently underdoped\cite{halp}.

For completeness and perspective, we finish this section by extracting
the information that follows from the CS equation (\ref{cs}). In order
to integrate this equation, we introduce the magnetization $\sigma =
\sigma(g)$ obtained by integrating the equation
    \begin{align}
      \begin{aligned}
	0 &= \left(\beta(g) \frac{\partial}{\partial g} +
	\frac{1}{2}\gamma(g)\right) \sigma(g)& \\
	\sigma(g_\mu) &= B \, \exp{\left\{ -\int^{g_\mu}
	\frac{\gamma(g)}{2\beta(g)} \, dg\right\}} \\
	&\simeq M \, \left[ 1 - \frac{g_\mu}{2 \pi^2} \right] = M
	\left( 1 - g_\mu / g_{UV}^c \right)
      \end{aligned}
    \end{align}
where $M$ is an arbitrary integration constant. Before continuing with
the analysis of the CS equation, we briefly digress to discuss the
properties of the coupling-dependent renormalized magnetization
$\sigma(g_\mu)$. For $g_\mu < 2 \pi^2 = g_{UV}^c$, i.e. in the N\'eel
ordered phase, the renormalized magnetization $\sigma(g_\mu)$ is a
nonzero constant. We see that as the coupling constant is tuned to the
critical value $g_{UV}^c$, the magnetization goes to zero with the
critical exponent $\beta=1$\footnote{This is not the $\beta$ function,
but this exponent is called $\beta$ for convention. We hope this does
not cause confusion.}. Since this is a one loop result, the exponent
is not accurate but the fact that the magnetization goes to zero at
criticality is true.

Let us go back to the analysis of the CS equation. In order to
simplify it, we define the correlation length via the equation
    \begin{align}
      \begin{aligned}
	0 &= \left(\mu \frac{\partial}{\partial\mu} + \beta(g)
	\frac{\partial}{\partial g}\right)\xi_J (\mu,g) \\
	\xi_J (g_\mu,\mu) &= \frac{1}{\mu} \exp{\left\{ \int^{g_\mu}
	\frac{dg'}{\beta(g')} \right\}} \\
	&= \frac{1}{\mu} \left[ \frac{g_\mu / g_{UV}^c}{1 -
	g_\mu/g_{UV}^c} \right]
      \end{aligned}
    \end{align}
This correlation length is the Josephson correlation length
$\xi_J$\cite{halp, sachdev3} which determines the crossover from short
distance critical behavior to long distance Goldstone behavior. Just
as with the magnetization, this determination of the correlation
length is not accurate enough because it is a one loop result. We see
that the correlation length diverges with the exponent $\nu = 1$.

\section{Topological Excitations of N\'eel Ordered $2+1$ D
 Antiferromagnets}

$2 + 1$ D antiferromagnets and their effective description via the
$O(3)$ nonlinear sigma model have a classical ``ground state'' or
lowest energy state with N\'eel order corresponding to a constant
magnetization. The equations of motion that follow from the action
have approximate time dependent solutions, corresponding to Goldstone
spin wave excitations. The equations of motion, in 2+1 D only, also
have exact static solitonic solutions of finite
energy\cite{polya1}. Digressing for a moment from antiferromagnets,
since the 1970's, it has been known that when systems have exact
classical, time independent solutions which are stable against quantum
fluctuations, these solutions are quantum particle excitations of the
system\cite{solitons}. The nonlinear sigma model, in $2 + 1$ D only,
possesses time independent solutions which are of a topological
nature\cite{gross1,polya1} and have finite energy. These excitations
are disordered at finite length scales but relax into the N\'eel state
far away
    \beq
      \lim_{|\vec x| \rightarrow \infty} \vec n=(0,0,1) \;, \qquad
      \qquad \lim_{|\vec x| \rightarrow \infty} w=0 \;.
    \eneq
They consist in the order parameter rotating a number of times as one
moves from infinity toward a fixed but arbitrary position in the
plane. Since two dimensional space can be thought of as an infinite 2
dimensional sphere, the excitations fall in homotopy classes of a 2D
sphere into a 2D sphere: $S^2 \rightarrow S^2$. The topological
excitations are thus defined by the number of times they map the 2D
sphere into itself. They are characterized by the Jacobian
    \beq
      q = \frac{1}{8 \pi} \int d^2 x \epsilon^{i j} \vec n \cdot
      \partial_i \vec n \times \partial_j \vec n 
    \eneq
or in terms of the stereographic variable
    \begin{align}
      \begin{aligned}
	q &= \frac{i}{2\pi} \epsilon^{i j} \, \int d^2 x \,
	\frac{\partial_i w \partial_j w^*}{ (1 + |w|^2)^2} \\
	&= \frac{1}{\pi} \int d^2 x \frac{\partial_z w \partial_{z^*}
	w^* - \partial_{z^*} w \partial_z w^*}{ (1 + |w|^2)^2} \;.
      \end{aligned}
    \end{align}
The number $q$ will be an integer measuring how many times
the $n$-sphere gets mapped into the infinite 2D sphere corresponding
to the plane where the spins live. If we define the space-time current
    \beq 
      J^\mu= \frac{1}{8 \pi} \epsilon^{\mu \nu \sigma } \vec n \cdot
      \partial_\nu \vec n \times \partial_\sigma \vec n =
      \frac{i}{2\pi} \, \epsilon^{\mu \nu \sigma} \,
      \frac{\partial_\nu w \partial_\sigma w^*}{(1 + |w|^2)^2} \; ,
    \eneq
it is easily seen that it is conserved $\partial_\mu J^\mu =
0$ and that the charge associated with it is our topological charge
    \beq
      q = \int d^2x \, J^0 \;.
    \eneq
Thus $q$ is a conserved quantum number. These topological field
configurations were originally discovered by Skyrme\cite{skyrme} and
are called skyrmions. The conserved charge is the skyrmion number.

Since the skyrmions are time independent, their energy is given by 
    \beq
      E_s = \frac{4 \Lambda}{g_\Lambda} \int d^2x \, \frac{ \partial_z
      w \partial_{z^*} w^* + \partial_{z^*} w \partial_z w^*}{\left( 1
      + |w|^2 \right)^2} \;.
    \eneq
From this expression and the one for the charge $q$, it is easily
seen\cite{gross1,polya1} that $E \ge 4 \pi |q| \Lambda/g_\Lambda$. We
see that we can construct skyrmions with $q > 0$ by imposing the
condition
    \beq 
      \partial_{z^*} w = 0
    \eneq
that is $w$ is a function of $z$ only. The magnetization, $\vec n$ or
$w$, is an analytic function of $z$ almost everywhere. The worst
singularities it can have are poles. The skyrmions will have a
location given by the positions of the poles or of the zeros of
$w$. Far away from its position, the field configuration will relax
back to the original N\'eel order. Therefore we have the boundary
condition $w(\infty) = 0$, which implies
    \beq
      w = \lambda^q \prod_{i=1}^q \frac{1}{z - a_i}
    \eneq
which can easily be checked to have charge $q$ and energy 
    \beq
      E_s (q) = 4 \pi \frac{\Lambda}{g_\Lambda} q = 4 \pi \rho_s q\;.
    \eneq
$\lambda^q$ is the arbitrary size and phase of the configuration and
$a_i$ are the positions of the skyrmions that constitute the
multiskyrmion configuration. The energy is independent of the the size
and phase due to the conformal invariance of the configuration. We
remark that since the multiskyrmions energy is the sum of individual
skyrmion energies, the skyrmions do not interact among
themselves\cite{gross1}. Similarly, the multiantiskyrmion
configuration can be shown to be
    \beq
      w = (\lambda^*)^q \prod_{i=1}^q \frac{1}{z^*-a_i^*}
    \eneq
with charge $-q$ and energy $4 \pi \Lambda q / g_\Lambda$. Skyrmions
and antiskyrmions do interact as shown in Appendix
\ref{topology}. In that Appendix we collect these results on
skyrmions and provide further developments.

We now investigate whether skyrmions and antiskyrmion configurations
are relevant at the quantum critical point. As mentioned above, their
classical energy is $4\pi\Lambda/g_\Lambda$, which is independent of
the size of the skyrmion $\lambda$. On the other hand, in real
physical systems there are quantum and thermal fluctuations. These
renormalize the effective coupling constant of the nonlinear sigma
model and make it scale dependent. To one loop order the renormalized
coupling constant is
    \beq
      \label{gren}
      g_\mu = \frac{\mu}{\Lambda} \frac{g_\Lambda}{1 -
      (g_\Lambda/2\pi^2)\left(1 - \mu/\Lambda\right)} \,.
    \eneq
Since the skyrmion has an effective size $\lambda$, spin waves of
wavelength smaller than $\lambda$ renormalize the energy of the
skyrmion via the coupling constant renormalization. The energy $E$ of
the skyrmion at the scale set by its size $\mu = 1/\lambda$ is now
    \beq
      E_s = \frac{4 \pi \mu}{g_\mu} \Big |_{\mu = 1/\lambda} =
      \frac{4 \pi \Lambda}{g_\Lambda} \left[ 1 - \frac{g_\Lambda}{2
      \pi^2} \left( 1 - \frac{1}{\lambda\Lambda} \right) \right] \;.
    \eneq
We see that the skyrmion energy now depends on its size through the
renormalization effects and thus the conformal invariance of the
configuration is broken. This is the well known phenomenon of broken
scale invariance in renormalizable theories due to the scale
dependence of the coupling constant\cite{cs1, dilation}. We see that
if we tune the system to criticality, $g_\Lambda = 2 \pi^2$, the
energy of a skyrmion of size $\lambda$ is $E_s = 2 / (\pi \lambda)$
and the energy of excitation for skyrmions of arbitrarily large size
is zero and hence degenerate with the ground state. The quantum
critical point thus seem to be associated with skyrmion gap
collapse. In order to see if skyrmion fluctuations are indeed relevant
to the critical point, let's be a little bit more careful and
explicit.

If the system is at temperature $T = 1/\beta$, this temperature sets
the size of the skyrmion to be the thermal wavelength $\lambda =
\beta$. The skyrmion Euclidean action is then
    \beq
      S_s = \frac{4\pi\beta}{\beta g_{1/\beta}} =
      \frac{4\pi\beta\Lambda}{g_\Lambda}\left[ 1 -
      \frac{g_\Lambda}{2\pi^2} \left( 1 - \frac{1}{\beta\Lambda}
      \right) \right]
    \eneq

The probability for skyrmion creation is given by $P \propto
e^{-S_s}$. Let us see how this probability behaves at low
temperatures. When in the N\'eel ordered phase, $g_\Lambda < g_c = 2
\pi^2$, the skyrmion Euclidean action diverges as $T \rightarrow
0$. Therefore, the probability for skyrmion contributions is
suppressed exponentially at low temperatures, vanishing at zero
temperature. {\it Skyrmions are gapped and hence irrelevant to low
temperature physics in the N\'eel ordered phase}.

At the quantum critical point $g_\Lambda = g_c = 2\pi^2$, the skyrmion
Euclidean action is
    \beq
      S_s = \frac{2}{\pi} \,.
    \eneq
This action is finite and constant at all temperatures and in
particular, it will have a nonzero limit as the temperature goes to 0:
the skyrmion probability is nonzero and constant at arbitrarily low
temperatures and zero temperature. {\it Hence there are skyrmion
excitations at criticality at arbitrarily low energies and
temperatures, including at zero temperature. Therefore skyrmion
excitations contribute to quantum critical physics}.

\section{Topologically Nontrivial Configurations with Zero Skyrmion 
Number}

We have seen that skyrmions are relevant at criticality as the
critical point is associated with skyrmion gap collapse and they have
a nonzero probability to be excited at arbitrarily low temperature at
criticality. On the other hand, skyrmions have nonzero conserved
topological number while the ground state has zero skyrmion
number. Absent any external sources that can couple directly to
skyrmion number, they will always be created in equal numbers of
skyrmions and antiskyrmions. Therefore, in order to study the effect
of skyrmions and antiskyrmions we need to include configurations with
equal number of skyrmions and antiskyrmions in the partition function.

We need to find nontrivial solutions of the nonlinear sigma model
equations of motion with zero skyrmion number that corresponds to
superpositions of equal number of skyrmions and antiskyrmions. The
classical equations of motion which follow by stationarity of the
classical action are
    \begin{align} 
      \begin{aligned} 
        \label{eom}
        \square w &= \frac{2 w^*}{1 + |w|^2}
	\partial^\mu w \partial_\mu w \\ 
	\partial_0^2 w - 4 \partial_z \partial_{z^*} w &=
	\frac{2 w^*}{1 + |w|^2} \left\{ (\partial_0 w)^2 - 4\partial_z
	w \partial_{z^*} w \right\}
      \end{aligned}
    \end{align}
We are interested in time independent and finite energy (or nonzero
probability) solutions. For time independent solutions, the structure
of the equations suggests a solution of the form
    \beq
      w = e^{i\varphi} \tan \left[ f(z) + \left( f(z) \right)^* +
      \frac{\theta}{2} \right]
    \eneq
where $\varphi$ and $\theta$ are arbitrary, constant angles and $f(z)$
is an arbitrary function of $z$ only and not of $z^*$, which is
analytic in $z$ almost everywhere (just as with skyrmion and
antiskyrmion solutions, the function can have poles but no worse
singularities). These solutions are topologically trivial since they
have zero skyrmion number. On the other hand, as we shall see
explicitly below, the finite energy solutions will correspond to
arbitrary superpositions of equal number of skyrmions and
antiskyrmions with $q = \pm n$. Since, despite being topologically
trivial, these solutions will be composed of topologically nontrivial
configurations (skyrmions and antiskyrmions), we dub them topolons. We
hope this name does not create confusion.

The first steps we need to check that the topolon solution written
above is a solution is to compute the derivatives. We start with the
simplest ones. Since our solutions are time independent, we
immediately obtain
    \beq
      \partial_0 w = 0 \;, \qquad \qquad \partial_0^2 w = 0 \,.
    \eneq
The next derivatives to calculate are
    \begin{align}
      \begin{aligned}
        \partial_z w &= e^{i\varphi} f'(z) \sec^2 \left[ f(z) + \left(
        f(z) \right)^* + \frac{\theta}{2} \right] \\
        \partial_{z^*} w &= e^{i\varphi} \left(f'(z)\right)^* \sec^2
        \left[ f(z) + \left( f(z) \right)^* + \frac{\theta}{2} \right]
        \\
        \partial_z \partial_{z^*} w &= 2 e^{i\varphi} |f'(z)|^2 \sec^2
        \left[ f(z) + \left( f(z) \right)^* + \frac{\theta}{2} \right]
        \\
        &\times \tan \left[ f(z) + \left( f(z) \right)^* +
        \frac{\theta}{2} \right] \,.
      \end{aligned}
    \end{align}
Using these derivatives, we see that the left hand side of the
equation of motion (\ref{eom}) is
    \begin{align}
      \begin{aligned}
	\text{LHS} &= -8 e^{i\varphi} |f'(z)|^2 \sec^2 \left[ f(z) +
	\left( f(z) \right)^* + \frac{\theta}{2} \right] \\
	&\times \tan \left[ f(z) + \left( f(z) \right)^* +
	\frac{\theta}{2} \right] \,.
      \end{aligned}
    \end{align}
and the right hand side is
    \begin{align}
      \begin{aligned}
	\text{RHS} &= \frac{2 e^{-i\varphi} \tan \left[ f(z) + \left(
	f(z) \right)^* + \frac{\theta}{2} \right]}{1 + \tan^2 \left[
	f(z) + \left( f(z) \right)^* + \frac{\theta}{2} \right]} \\
	&\left\{ - 4 e^{2i\varphi} |f'(z)|^2 \sec^4 \left[ f(z) +
	\left( f(z) \right)^* + \frac{\theta}{2} \right] \right\} \\
	&= -8 e^{i\varphi} |f'(z)|^2 \sec^2 \left[ f(z) + \left( f(z)
	\right)^* + \frac{\theta}{2} \right] \\
	&\times \tan \left[ f(z) + \left( f(z) \right)^* +
        \frac{\theta}{2} \right] \;.
      \end{aligned}
    \end{align}
The left hand side and the right hand side are identical and therefore
the topolon solution written above is indeed an exact solution. 

The energy of this solution is easily calculated to be
    \beq E =
      \frac{8\Lambda}{g_\Lambda} \int d^2x \Big | \frac{df(z)}{dz}
      \Big |^2 \,.
    \eneq
In order for the solution to have finite energy, $f(z)$ must decay
faster than $1/z^\xi$ as $z \rightarrow \infty$, with $\xi>0$. But
since $f(z)$ is analytic except for poles or multipoles, $f(z)$ decays
at least as fast as $1/z$ as $z \rightarrow \infty$. Therefore, a
general form that has finite energy is given by
    \beq
      f_n(z) = \lambda^n \prod_{i=1}^n
      \frac{1}{z-a_i}
    \eneq
with $w_n$ given as above with $f(z) = f_n(z)$. This is the general
$n$-topolon. Two parameters the $n$-topolon depends on are the two
orientation angles. In general, skyrmions depend on two orientation
angles which are given by the orientation of the N\'eel order the
skyrmion relaxes on far away. These are usually not counted when
counting the parameters of a skyrmion as they are usually fixed to
constant values by the boundary conditions. Besides these two angles,
a skyrmion depends on an arbitrary complex parameter $\lambda$ and $n$
complex positions when we have an $n$-skyrmion. The $n$-topolon
depends also on an arbitrary complex parameter $\lambda$ and $n$
complex positions. Besides the arbitrariness of these parameters (the
same arbitrariness as for an $n$-skyrmion), there is no other
arbitrariness to the topolon as its form is dictated by solving the
equations of motion. 

Given that the argument of the topolon is precisely the sum of an
$n$-skyrmion with an $n$-antiskyrmion, it is clear that a topolon is
in general a superposition of skyrmions and antiskyrmions. The
antiskyrmions are at the same positions as the skyrmions, i.e. the
argument of the topolon has paired skyrmions and antiskyrmions at the
same position. If this was not the case, we would not have a solution
of the equations of motion. This is perhaps not surprising because if
we take a look at the skyrmion-antiskyrmion interaction we obtained in
Appendix \ref{topology}, it has a minimum when the skyrmion and
antiskyrmion are at the same position, that is, the relative distance
is zero.

\section{Renormalization of the Nonlinear Sigma Model Including 
Goldstone and Topolons}

We have just found time independent solutions to the equations of
motion given by
    \beq
      w_t^{(n)} = e^{i\varphi} \tan \left[ \prod_{i=1}^n
      \frac{\lambda}{z - a_i} + \prod_{i=1}^n \frac{\lambda^*}{z^* -
      a_i^*} + \frac{\theta}{2} \right]
    \eneq
which corresponds to superpositions of equal number of skyrmions and
antiskyrmions. These solutions contribute to the partition function
for the antiferromagnet, which is then given by
    \beq
      \mathcal Z = \sum_{n=0}^\infty \mathcal Z_n
    \eneq
where
    \beq
      \mathcal Z_0 = \int \frac{\mathcal D \nu \mathcal D\nu^*}{(1 +
      |\nu|^2)^2} e^{-S [\nu]}
    \eneq
is the usual partition function for the nonlinear sigma model with no
topolons and only the spin wave like fields $\nu$, and $S$ is the
Euclidean action for the nonlinear sigma model in terms of the
stereographic projection variable $w = \nu$. We also have that
    \begin{align}
      \begin{aligned}
	\mathcal Z_{n\neq 0} &= \int \frac{\mathcal D\nu \mathcal D
	\nu^*}{(1 + | w_t^{(n)} + \nu|^2)^2} \frac{d\Omega}{4\pi}
	\frac{d\lambda}{1/\Lambda} \\
	&\times \prod_{i = 1}^n \frac{d^2 a_i}{A} \; e^{-S [w_t^{(n)}
	+ \nu]} \,.
      \end{aligned}
    \end{align}
The $Z_{n\neq 0}$ is the path integral with the $n$ topolons with spin
waves $\nu$. Besides integrating over the spin wave configurations, we
must integrate over the topolon parameters: its size $\lambda$
normalized to the lattice spacing $1/\Lambda$, the positions $a_i$ of
the topolon constituents skyrmions and antiskyrmions, normalized to
the area $A$ of the system, and over the solid angle of the topolon
orientation normalized to $4\pi$.

The next step to evaluate the partition function, or to evaluate
expectation values from it, would be to expand the topolon action in a
semiclassical expansion about the spin waves. This process can be
simplified by realizing that the same physics follows if all the
constituents skyrmions are placed in the same position. This result is
not obvious so we will review the main points of how it comes
about. When the semiclassical expansion is performed, as will be
described below, we obtain a sum of terms in the partition function,
each of which is weighted by $e^{-\beta E_n}$ where $E_n$ is the
$n$-topolon energy. This energy is given by
    \beq
      E_n = \frac{8\pi\Lambda n}{g_\Lambda} \left[
      \frac{\lambda}{R(a_1,a_2,\dots,a_n)} \right]^{2n}
    \eneq
where $R$ is a distance that depends on the particular positions of
the skyrmions and antiskyrmions making up the arguments of the
$n$-topolon. All terms in the semiclassical expansion, after being
multiplied by the weight factor, need to be integrated over the solid
angle of the orientation of the topolons ($d\Omega/4\pi$) and over the
positions of the $n$ skyrmions and antiskyrmions that constitute the
argument of the $n$-topolon ($d^2 a_1 \, d^2 a_2\, \dots \, , d^2 a_n
/ (\text{area})^n$). The only dependence on the $a_i$'s in the
weighting factor appears through $R(a_1,\dots,a_n)$. We also need to
integrate over $\lambda$. Integration over solid angles of each of the
terms of the semiclassical expansion yields a constant independent of
$a_i$'s. That is, the angular average over powers of $w_t^{(n)}$ is
independent of the positions $a_i$'s. 

We make a change of variables and interchange the variable $a_1$ with
the variable $R$. When we do this, we are left with integrals with
volume element
    \beq
      \frac{\partial a_1}{\partial R} \, d\lambda dR \prod_{i=2}^n
      da_i \,.
    \eneq
We then perform a change of variables in the $\lambda-R$ plane:
    \beq
      \lambda' = \frac{\lambda}{R} \; , \qquad \qquad R' = R \,.
    \eneq
The volume element of our integral becomes
    \beq
      R' \, \frac{\partial a_1}{\partial R'} \, d\lambda' dR'
      \prod_{i=2}^n da_i \,.
    \eneq
Now the weighting factor $e^{-\beta E_n}$ depends on $\lambda'$ only
and is independent of $R'$ and the $a_i$'s. Hence the integrations
over $R'$ and the $a_i$'s can be performed yielding an irrelevant
constant factor which can be absorbed in the normalization or
dropped. We are left with an integration over $\lambda'$ and a
weighting factor which is equal to the one obtained with the $n$
skyrmions and antiskyrmions that constitute the $n$-topolon placed in
the same position 
    \beq
      w_t^{(n)} = e^{i \varphi} \tan \left[ \left( \frac{\lambda}{z -
      a} \right)^n + \left( \frac{\lambda^*}{z^* - a^*} \right)^n +
      \frac{\theta}{2} \right]
    \eneq
since the Euclidean action of such a configuration is given by
    \begin{align}
      \begin{aligned}
        &S_t = \beta E_n \\
	&= \frac{4 \beta \mu}{g_\mu} \int d^2x \frac{\partial_z
        w_t^{(n)} \, \partial_{z^*} w_t^{(n)*} + \partial_{z^*}
        w_t^{(n)} \, \partial_z w_t^{(n)*}}{ (1 + |w_t^{(n)}|^2)^2} \\
	&= \frac{8 \beta \mu}{g_\mu} \, n^2 \lambda^{2n} \int d^2x
	\frac{1}{|z|^{2(n + 1)}} = \frac{8 \pi \beta \mu}{g_\mu} \, n
	\left (\lambda \Lambda \right)^{2n} \;.
      \end{aligned}
    \end{align}
For the topolon the scale is set by the size of the topolon, so we
choose $\mu = 1/\lambda$. In the case where $\lambda$ is set by the
temperature ($\beta = \lambda$), we have
    \begin{align}
      \begin{aligned}
	S_t &= \beta E_n = \int_0^\lambda d\tau \frac{8 \pi}{\lambda
	g_{1/\lambda}} \, n \left( \lambda \Lambda \right)^{2n} \\
	&\simeq \frac{8 \pi \lambda \Lambda}{g_\Lambda} \, n \left(
	\lambda \Lambda \right)^{2n} \;.
      \end{aligned}
    \end{align}

Without loss of generality, from now on we only consider the topolon
with all skyrmions constituting $f(z)$ placed at the same position
$a$. For this case we have
    \begin{align}
      \begin{aligned}
	&\mathcal Z_{n\neq 0} = \\
	&\int \frac{\mathcal D\nu \mathcal D \nu^*}{(1 + | w_t^{(n)} +
	\nu|^2)^2} \frac{d^2a}{A} \frac{d\Omega}{4\pi}
	\frac{d\lambda}{1/\Lambda} \; e^{-S[w_t^{(n)} + \nu]} \,.
      \end{aligned}
    \end{align}
For convenience computing averages, we will normalize the partition
somewhat differently, so that we have
    \beq
      \label{fullpart}
      \mathcal Z = \sum_{n=0}^{\infty} \int \frac{\mathcal D\nu \,
      \mathcal D\nu^* \, d\lambda \, d^2 a \, d\varphi \, d\theta \sin
      \theta}{(1 + |w_t^{(n)} + \nu|^2)^2} \; \frac{e^{-S(w_t^{(n)} +
      \nu)}}{\mathcal Z_0}
    \eneq
The weight in the partition function is the appropriate nonlinear
sigma model Euclidean action
    \begin{align}
      \begin{aligned}
	&S(w_t^{(n)} + \nu) = \\
	&\frac{2\Lambda}{g_\Lambda} \int d^3x \; \frac{\partial_\mu
	(w_t^{(n)*} + \nu^*)\partial_\mu (w_t^{(n)} + \nu)}{(1 +
	|w_t^{(n)} + \nu|^2)^2} \;.
      \end{aligned}
    \end{align}
$\mathcal Z_0$ is an arbitrary normalization factor since multiplying
the partition function by a constant does not change the physics. We
will choose $\mathcal Z_0$ conveniently to make some of our
intermediate calculations look simpler.

Of course, the partition function (\ref{fullpart}) cannot be
calculated exactly and we will approximate it by doing perturbation
theory about the noninteracting action
    \begin{align}
      \begin{aligned}
	S_0 &= \frac{2\Lambda}{g_\Lambda} \int d\tau d^2x \; \left[
	\frac{\partial_\mu w_t^{(n)*} \partial_\mu w_t^{(n)}}{(1 +
	|w_t^{(n)}|^2)^2} + \partial_\mu \nu^*\partial_\mu \nu \right]
	\\
	&= S_t + S_\nu
      \end{aligned}
    \end{align}
where the first action is the unperturbed topolon action and the
second is a free spin wave action neglecting the nonlinearity which
accounts for the interactions. Then the partition function is
    \begin{align}
      \begin{aligned}
	\mathcal Z = \frac{1}{\mathcal Z_0} \sum_{n=0}^{\infty} &\int
	\frac{\mathcal D\nu \, \mathcal D\nu^* \, d\lambda \, d^2a \,
	  d\varphi \, d\theta \sin \theta}{(1 + |w_t^{(n)} + \nu|^2)^2}
	\; e^{-S_0 - S_I} \\
	S_I &= S(w_t^{(n)} + \nu) - S_0 \;.
      \end{aligned}
    \end{align}
We choose
    \beq
      \mathcal Z_0 = \sum_{n=0}^{\infty} \int \frac{\mathcal D\nu \,
      \mathcal D\nu^* d\lambda \, d^2 a \, d\varphi \, d\theta \, \sin
      \theta}{(1 + |w_t^{(n)}|^2)^2} \; e^{-S_0} \;.
    \eneq

We perform perturbation theory in the interaction term by integrating
out short distance degrees of freedom. This is done by integrating
large momentum and frequency spin waves, those with values between the
microscopic cutoff $\Lambda$ and a lower renormalization scale
$\mu$. In order for perturbation theory to be controlled we keep $\mu$
close to $\Lambda$ and improve this perturbation theory by using the
renormalization group\cite{weinberg, br1, br2, zj}. To integrate the
short distance topolon configurations we integrate those with sizes
between the lattice constant $1/\Lambda$ and the renormalization
distance $1/\mu$. Since $\mu$ is close to $\Lambda$ we expand the
action to lowest order in the interaction term and a simple
manipulation leads to a partition function with a lower cutoff $\mu$
and with effective action correct to first order given by
    \beq
      S_{\text{eff}}^\mu = S_0 + \langle S_I \rangle \;.
    \eneq
The average $\langle S_I\rangle$ can be calculated by expanding the
interacting action to fourth order in $\nu$ and $w_t$ as higher order
terms in the expansion contribute higher orders in $\mu - \Lambda$ and
are suppressed for $\mu$ very close to $\Lambda$. This expansion
yields the 13 terms
    \begin{align}
      \begin{aligned}
	I &= \frac{2\Lambda}{g_\Lambda} \int d^3 x \langle
	\partial_\mu w_t^n\partial_\mu w_t^{n*}\rangle \\
	II &= -\frac{4\Lambda}{g_\Lambda} \int d^3 x \langle
	\partial_\mu w_t^n\partial_\mu w_t^{n*} |w_t^n|^2 \rangle \\
	III &= -\frac{4\Lambda}{g_\Lambda} \int d^3 x \langle
	\partial_\mu w_t^n\partial_\mu w_t^{n*} |\nu|^2 \rangle \\
	IV &= -\frac{4\Lambda}{g_\Lambda} \int d^3 \langle
	\partial_\mu w_t^n\partial_\mu w_t^{n*} w_t^n\nu^* \rangle +
	C.C. \\
	V &= \frac{2\Lambda}{g_\Lambda} d^3 x \langle \partial_\mu\nu
	\partial_\mu w_t^{n*} \rangle + C.C. \\
	VI &= -\frac{4\Lambda}{g_\Lambda} \int d^3 x \langle
	\partial_\mu\nu \partial_\mu w_t^{n*} |w_t^n|^2 \rangle +
	C.C. \\
	VII &= -\frac{4\Lambda}{g_\Lambda} \int d^3 x \langle
	\partial_\mu \nu \partial_\mu w_t^{n*} |\nu|^2 \rangle +
	C.C. \\
	VIII &= -\frac{4\Lambda}{g_\Lambda} \int d^3 x \langle
	\partial_\mu \nu \partial_\mu w_t^{n*} w_t^n\nu^* \rangle +
	C.C. \\
	IX &= -\frac{4\Lambda}{g_\Lambda} \int d^3x \langle
	\partial_\mu\nu \partial_\mu w_t^{n*} w_t^{n*}\nu\rangle +
	C.C. \\
	X &= \frac{2\Lambda}{g_\Lambda} d^3x \langle \partial_\mu \nu
	\partial_\mu \nu^* \rangle \\
	XI &= -\frac{4\Lambda}{g_\Lambda} \int d^3x \langle
	\partial_\mu \nu \partial_\mu \nu^* |w_t^{n*}|^2 \rangle \\
	XII &= -\frac{4\Lambda}{g_\Lambda} \int d^3x \langle
	\partial_\mu \nu \partial_\mu \nu^* |\nu|^2\rangle \\
	XIII &= -\frac{4\Lambda}{g_\Lambda} \int d^3x \langle
	\partial_\mu \nu \partial_\mu \nu^* w_t^n \nu^* \rangle + C.C.
      \end{aligned}
    \end{align}

When we perform the averages, all terms with space derivatives of
$w_t$ are irrelevant as the derivatives can be interchanged by a
derivative over the position coordinates of the topolons and when
averaged, will be smaller than the other terms by a factor of the
lattice constant divided by the linear dimensions of the system, a
ratio that goes to zero in the thermodynamic limit. Hence we are left
only with the terms $X$ to $XIII$. Term $XIII$ being odd in the spin
waves averages to zero. The average of term $X$ over short distance
gives a constant and is thus irrelevant. We are left with the
averaging terms $XI$ and $XII$. If we go to higher powers in the
expansion, there will be similar terms to $XI$ with higher powers of
$|w_t^{(n)}|^2$. These can all be summed. We then obtain that the
effective action after averaging over short distance fluctuations is
    \begin{align}
      \begin{aligned}
	&-\frac{4 \Lambda}{g_\Lambda} \int d^3x \, \partial_\mu \nu \,
	\partial_\mu \nu^* \langle |\nu|^2 \rangle \\
	&+ \frac{2 \Lambda}{g_\Lambda} \int d^3x \, \partial_\mu \nu \,
	\partial_\mu \nu^* \left \langle \frac{1}{( 1 + |w_t^{(n)}|^2
	)^2} \right \rangle \;.
      \end{aligned}
    \end{align}
The first term, when averaged over short distance fluctuations gives
the already calculated Goldstone renormalization
    \beq
      \frac{- 4 \Lambda}{g_\Lambda} \frac{g_\Lambda}{4\pi^2} \left[ 1
      - \frac{\mu}{\Lambda} \right] \int d^3x \, \partial_\mu \nu \,
      \partial_\mu \nu^* \;.
    \eneq
To average over short distance topolons, we first perform the angular
average and the average over positions, which gives 
    \beq
      \left \langle \frac{1}{( 1 + |w_t^{(n)}|^2 )^2} \right
      \rangle_{\text{angles} + \text{positions}} = \frac{1}{3} \;.
    \eneq
For this average we do not include the weighting factor, as this
factor is independent of the angles and positions. We now integrate
over topolon sizes $\lambda$ from $1/\Lambda$ to $1/\mu$, including
the weighting factor to finally obtain
    \beq
      \frac{2 \Lambda}{3 g_\Lambda} \left[ 1 - \frac{\mu}{\Lambda}
      \right] \exp \left[ - n \frac{8 \pi}{g_\Lambda} \right] \int
      d^3x \, \partial_\mu \nu \, \partial_\mu \nu^* \;.
    \eneq
Since there are topolon contributions from $n = 1$ to $n = \infty$, we
sum these contributions from 1 to $\infty$ to obtain
    \begin{align}
      \begin{aligned}
        \frac{2 \Lambda}{3 g_\Lambda} &\left[ 1 - \frac{\mu}{\Lambda}
	\right] \sum_{n = 1}^\infty \exp \left[ -n \frac{8
	\pi}{g_\Lambda} \right] \int d^3x \, \partial_\mu \nu \,
	\partial_\mu \nu^* \\
	= &\frac{2 \Lambda}{3 g_\Lambda} \left[ 1 -
	\frac{\mu}{\Lambda} \right] \frac{1}{e^{8 \pi / g_\Lambda} -
	1} \int d^3x \, \partial_\mu \nu \, \partial_\mu \nu^* \;.
      \end{aligned}
    \end{align}
Both Goldstone and topolon renormalizations are proportional to $1 -
\mu / \Lambda$, which can be made as small as possible to make sure the 
perturbation expansion is controlled. The Goldstone renormalization is
linear in the coupling constant, but the topolon contribution is not
analytic in the coupling constant around $g_\Lambda = 0$. It is thus
non-perturbative. The renormalized action, renormalized spin stiffness
and beta function are
    \begin{widetext}
      \begin{align}
	\begin{aligned}
	  &S_{\text{ren}} = \frac{2\mu}{g_\mu} \int d^3x
	  \frac{\partial_\mu\nu \partial_\mu\nu^*}{(1 + |\nu|^2)^2}
	  \simeq \frac{2\Lambda}{g_\Lambda} \int d^3x \partial_\mu\nu
	  \partial_\mu\nu^* \left\{ 1 - \left(1 -
	  \frac{\mu}{\Lambda}\right) \frac{g_\Lambda}{2\pi^2} + \left(
	  1 - \frac{\mu}{\Lambda} \right)
	  \frac{1}{3(e^{8\pi/g_\Lambda} - 1)} \right\} \\
	  &\rho_s(\mu) = \frac{\mu}{g_\mu} = \frac{\Lambda}{g_\Lambda}
	  \left\{ 1 - \left(1 - \frac{\mu}{\Lambda}\right)
	  \frac{g_\Lambda}{2\pi^2} + \left( 1 - \frac{\mu}{\Lambda}
	  \right) \frac{1}{3(e^{8\pi/g_\Lambda} - 1)} \right\} \\
	  &\beta (g) = \mu \frac{\partial g}{\partial\mu}
	  \Big|_{\Lambda = \mu} = g - \frac{g^2}{2\pi^2} +
	  \frac{g}{3(e^{8\pi/g} - 1)} \;.
	\end{aligned}
      \end{align}
    \end{widetext}
The last term in the spin stiffness and in the beta function is the
contribution from the topolons and the rest is the contribution of the
spin waves to fourth order in the coupling constant. The coupling
constant at the quantum critical point $g_c\simeq 23.0764$ is obtained
from $\beta(g_c) = 0$.

The important point is that the topolons add a new term to the beta
function calculated above. This will lead to modification of the
critical properties beyond those calculated from the spin wave
expansion only. Below we estimate exponents from our
approxiamtions. These exponents will not be suffiently accurate for
two reasons. First, we calculated our spin wave expansion to first
order in the coupling constant $g$. This will lead to large
inaccuracies as $g$ is not small. On the other hand, this would be
easy to improve by including higher order corrections obtainable from
the $d = 4 - \epsilon$ expansion, which is Borel summable and thus
very accurate as far as the spin wave effects
concern\cite{zj}. Inclusion of these terms will not change the new
term we found from the topolons and hence will not affect the fact
that there is new critical physics coming from superpositions of
skyrmions and antiskyrmions. On the other hand such corrections are
necessary for accurate enough exponents.

Another issue is that there might be more exact solutions that
correspond to superpositions of equal numbers of skyrmions and
antikyrmions. We found no other solutions, but it is expected that
there should be solutions that are time dependent in which the
skyrmions and antiskyrmions are in motion. These need to be included
in order to have the most accurate possible exponents, but they do not
invalidate the physics we have found: that skyrmions and antiskyrmions
modify critical properties. They further add to the physics we have
found in the present work. Finally, for completeness we compare below
with other approaches that people use to calcualte exponents. Such
approaches are the $1/N$ expansion and the $d = 2 + \epsilon$
expansion. These are not as accurate as the $d = 4 - \epsilon$
expansion, which seems to give the accepted Heisenberg
exponents\cite{zj}, but this last expansion cannot capture topological
contributions as these exist only in two spatial dimensions.

The correlation length at scale $\mu$ is given by\cite{polya2}
    \beq
      \xi \sim \frac{1}{\mu} \exp \left[ \int_{g_c}^{g_\mu}
      \frac{dg}{\beta (g)} \right] \sim \left( g_c - g_\mu
      \right)^{1/\beta '(g_c)} \,.
    \eneq
The correlation length exponent is $\nu = -1/\beta ' (g_c) \equiv
-(d\beta/dg|_{g=g_c})^{-1}$. Including topolon contributions, it
evaluates to $\nu=0.9297$. The $d=2+\epsilon$ expansion of the $O(N)$
vector model, which agrees with the $1/N$ expansion for large $N$,
gives $\nu=0.5$\cite{zj}. We note that our value is larger than the
accepted numerical evaluations of critical exponents in the Heisenberg
model, $\nu=0.71125$\cite{nu}, but about as close to this accepted
Heisenberg value than the $2+\epsilon$ expansion or the $1/N$
expansion. We conjecture that the difference between our value and the
Heisenberg value is real and attributable to quantum critical degrees
of freedom.

Goldstone renormalizations of the ordering direction $\sigma = n_3$,
and hence of the anomalous dimension $\eta$, are notoriously
inaccurate. The one loop approximation leads to a value of $\eta=2$,
thousands of percent different from the accepted numerical Heisenberg
value of $\eta\simeq 0.0375$\cite{nu}. The large $N$ approximation,
which sums bubble diagrams, is a lot more accurate. To order $1/N$ one
obtains $\eta = 8/(3\pi^2 N) \simeq 0.09$ for $N=3$. We now calculate
the value of $\eta$ from topological nontrivial configurations
    \begin{align}
      \begin{aligned}
        &\langle n_3^2 \rangle = Z = 1 - \langle n_1^2 + n_2^2 \rangle
        \\
	&= 1 - \left\langle \frac{4 |w|^2}{(1 + |w|^2)^2}
	\right\rangle \simeq 1 - \frac{2}{3} \frac{1 - \mu/ \Lambda}{
	e^{8\pi/g_\Lambda} - 1 } \\
	&\Rightarrow \eta (g) = \frac{\mu}{Z}\frac{\partial
	Z}{\partial\mu}\Big |_{\Lambda = \mu} \simeq
	\frac{2}{3(e^{8\pi/g} - 1)} \;.
      \end{aligned}
    \end{align}
For the anomalous dimension at the quantum critical point we obtain
$\eta(g_c) = 0.3381$. On the other hand, we have seen that spin wave
contributions tend to give quite large and nonsensical values of
$\eta$. In fact so large as to wash out the momentum dependence of the
propagator. Hence, to calculate $\eta$, spin wave contributions prove
to be tough to control. Our calculation gives a value quite larger
than the accepted numerical value.  We have recently calculated
\cite{us} the unique value of $\eta$ that follows from quantum
critical fractionalization into spinons and find $\eta = 1$. While our
value obtained from topolons is far from 1, it is a lot closer than
the accepted numerical Heisenberg value and the $1/N$ value.

The main purpose of this work is to look for possible new physics due
to intrinsic critical excitations of quantum critical points. In this
article we studied the approach to the quantum critical point from the
N\'eel ordered phase. We have uncovered some things that were not
known before. One thing which has been hinted at before was that
skyrmion excitations are relevant to the quantum critical point. We
have provided strong evidence that skyrmion fluctuations contribute to
the quantum critical point. For the first time, we found new exact
solutions that corresponds to superpositions of equal number of
skyrmions and antiskyrmions. Also for the first time, we developed and
calculated the inclusion of these topological effects in the
renormalization group to yield critical exponents.

\appendix

\section{Stereography and Topology}
\label{topology}

\subsection{$2+1$ D Antiferromagnets}

In the present work we study 2 dimensional $O(3)$ quantum
antiferromagnets on bipartite lattices as described by the the
Heisenberg Hamiltonian
    \beq
      H = J \sum_{\langle i j\rangle} \vec S_i \cdot \vec S_j
    \eneq
with $J>0$ and $\langle i j\rangle$ meaning that $i j$ are next
neighbors. Haldane\cite{hal1} showed that in the large $S$ limit, the
low energy universal physics of the Heisenberg antiferromagnet is
equivalent to that given by the $O(3)$ nonlinear sigma model described
by the Lagrangian and action
    \beq
      L = \frac{\Lambda}{2 g} \int \, d^2 x \, \eta^{\mu\nu}
      \partial_\mu \vec n \cdot \partial_\nu \vec n = \frac{\Lambda}{2
      g} \int \, d^2 x \, \partial^\mu \vec n \cdot \partial_\mu \vec
      n
    \eneq
    \beq
      S = \int \, dt \,   L \;,
    \eneq
where $\eta^{\mu\nu}$ is the $2+1$ Lorentz metric and $\vec n$ is a 3
dimensional unit vector, $\vec n \cdot \vec n =1$, that represents the
sublattice magnetization. Physically the inverse coupling constant is
proportional to the ``spin'' or magnetization stiffness $\rho_s$.  The
large $S$ identification of the Heisenberg and nonlinear sigma model
holds because the amplitude fluctuations of the spins are irrelevant
for large $S$\cite{hal1}. Hence as long as the amplitude fluctuations
are irrelevant to the long distance physics, the $O(3)$ nonlinear
sigma model will be an apt description of antiferromagnets {\it
regardless of $S$}. One expects amplitude fluctuations to be less
relevant for lower dimensionality.
 
A very useful way of describing the $O(3)$ nonlinear sigma model is
through the stereographic projection\cite{gross1}: 
    \beq 
      n^1 \!+ i n^2 \!= \!\frac{2 w}{|w|^2 \!+ \!1} \;, \; n^3 \!=
      \!\frac{1 \!- \!|w|^2}{1 \!+ \!|w|^2} \;, \; w \!= \!\frac{n^1
      \!+ \!i n^2}{1 \!+ \!n^3} \;.
    \eneq 
The stereographic projection maps the sphere subtended by the N\'eel
field or staggered magnetization $\vec n$ to the complex $w$ plane by
placing the N\'eel sphere below the $w$ plane, with the north pole
touching the center of the plane. If one draws a straight line joining
the south pole and the point on the sphere corresponding to the N\'eel
field, then the mapping to the $w$ plane of this N\'eel direction is
given by extending the mentioned line until it intersects the $w$
plane.

In terms of $w$ the Lagrangian is 
    \begin{align}
      L &= \frac{2 \Lambda}{g}\int d^2x \frac{\partial^\mu w
      \partial_\mu w^*}{ (1 + |w|^2)^2} \\
      \nonumber
      &= \frac{2 \Lambda}{g} \int d^2x \frac{\partial_0 w \partial_0
      w^* - 2 \partial_z w \partial_{z^*} w^* - 2 \partial_{z^*} w
      \partial_z w^*}{(1 + |w|^2)^2} \;,
    \end{align}
where $z = x + i y$ and $z^* = x -i y$ is its conjugate. The classical
equations of motion which follow by stationarity of the classical
action are $\Box \vec n = 0$, which in terms of the stereographic
variable $w$ are
    \begin{align}
      \begin{aligned}
        \Box w =& \frac{2 w^*}{1 + |w|^2}\partial^\mu w \partial_\mu w
        \text{ or} \\
        \partial_0^2 w \!- 4\partial_z \partial_{z^*} w =& \frac{2
        w^*}{1 + |w|^2} \left[ (\partial_0 w)^2 \!- 4 \partial_z w
        \partial_{z^*} w \right] \;.
      \end{aligned}
    \end{align}

The quantum mechanics of the $O(3)$ nonlinear sigma model is achieved
either via path integral or canonical quantization. The last is
performed by defining the momentum conjugate to $\vec n$, or to $w$
and $w^*$, by
    \begin{align}
      \begin{aligned}
        \vec \Pi(t,\vec x) &\equiv \frac{\delta L}{\delta
        \partial_0\vec n(t,\vec x)} \\
	\Pi^*(t,\vec x) \equiv \frac{\delta L}{\delta \partial_0
	  w(t,\vec x)} \, &, \quad \Pi(t,\vec x) \equiv \frac{\delta
	  L}{\delta \partial_0 w^*(t,\vec x)}\;.
      \end{aligned}
    \end{align}
and then imposing canonical commutation relations among the momenta
and coordinates. Due to the nonlinear constraint $\vec n \cdot
\vec n =1$, the momentum $\vec \Pi$ is an angular momentum satisfying
the $SU(2)$ algebra: 
    \beq
      \vec \Pi \cdot \vec n = 0 \, , \quad \vec \Pi \times \vec \Pi =
      i \vec \Pi \; .
    \eneq
The Hamiltonian is then given by 
    \begin{align}
      \begin{aligned}
        &H = \int d^2x \left( \vec \Pi \cdot \partial_0 \vec n - L
	\right) \\
	&= \int d^2x \left[ \frac{g}{2 \Lambda} \vec \Pi^2 +
	\frac{\Lambda}{2 g} \partial_i \vec n \cdot \partial_i \vec n
	\right] \\
	&= \int d^2x \left( \Pi^* \cdot \partial_0 w + \Pi \cdot
	\partial_0 w^* - L\right) \\
	&= \int d^2x \left[ \frac{g}{2\Lambda} (1 + |w|^2)^2 \Pi^* \Pi
	+ \frac{2 \Lambda}{g} \frac{\partial_i w \partial_i w^*}{(1 +
	|w|^2)^2}\right] \\
	&= \int d^2x \left[ \frac{g}{2 \Lambda}(1 + |w|^2)^2 \Pi^* \Pi
	\right. \\
	&+ \left. \frac{4 \Lambda}{g} \frac{( \partial_z w
	\partial_{z^*} w^* \partial_{z^*} w \partial_z w^*)}{(1 +
	|w|^2)^2} \right] \;.
      \end{aligned}
    \end{align}

The Heisenberg equations of motion that follow from this Hamiltonian,
when properly ordered, are identical to the classical equations. There
are ordering ambiguities in this Hamiltonian. The usual prescription
to deal with the ambiguities is by symmetrization, but the correct
order can only be determined by comparison with experiment if there is
a measurement that is sensitive to the operator order. Most results
are insensitive to these ordering ambiguities as they only introduce
short distance modifications to the physics.

\subsection{Excitations of the N\'eel Ordered Phase of $O(3)$ 
Nonlinear Sigma Model}

We remind the reader that classically the lowest energy state is
N\'eel ordered for all $g < \infty$, i.e. the spin stiffness,
$\rho_s$, is never zero. Quantum mechanically the situation is
different. In $2+1$ and higher dimensions, quantum mechanical
fluctuations cannot destroy the N\'eel order for the bare coupling
constant less than some critical value $g_c$\cite{polya2,br1,br2}. At
$g_c$ the renormalized long-distance, low-energy coupling constant
diverges\cite{polya2}, i.e. the system loses all spin stiffness. At
such a point quantum fluctuations destroy the N\'eel order in the
ground state as the renormalized stiffness vanishes. In the present
section we concentrate in the excitations of the N\'eel ordered phase.

\subsection{Magnons}

Linearization of the equations of motion leads to the low energy
excitations of the sigma model (magnons in the N\'eel phase and
triplons in the disordered phase) when quantized. We now turn our
attention to the N\'eel ordered phase. When the system N\'eel orders,
$\vec n$, or equivalently $w$, will acquire an expectation value:
    \beq
      \langle n^a \rangle = -\delta^{3a} \, , \quad \left \langle
      \frac{1}{w} \right \rangle = 0 \;.
    \eneq
where we have chosen the order parameter in the $-3-$direction as it
will always point in an arbitrary, but fixed direction. Small
fluctuations about the order parameter
    \beq
      \frac{1}{w} = \nu
    \eneq
are the magnons or Goldstone excitations of the N\'eel phase. To
leading order the magnon Lagrangian is
    \begin{align}
      \begin{aligned}
        &L = \frac{2 \Lambda}{g} \int d^2x \frac{\partial^\mu \nu
        \partial_\mu \nu^*}{ (1 + |\nu|^2)^2} \simeq \frac{2
        \Lambda}{g} \int d^2x \, \partial^\mu \nu \partial_\mu \nu^*
        \\
	&\!= \frac{2 \Lambda}{g} \! \int d^2x \left( \partial_0 \nu
	\partial_0 \nu^* \!- 2 \partial_z \nu \partial_{z^*} \nu^* \!-
	2 \partial_{z^*} \nu \partial_z \nu^* \right)
      \end{aligned}
    \end{align}
leading to the equations of motion
    \beq
      \Box \nu=0 \;, \qquad \partial_0^2 \nu - 4 \partial_z
      \partial_{z^*} \nu =0 \; .
    \eneq
The linearized excitations of the N\'eel phase have relativistic
dispersion that vanishes at long wavelengths as dictated by
Goldstone's theorem\cite{gold}. The magnons are of course spin $1$
particles. They have only 2 polarizations as they are transverse to
the N\'eel order.

\subsection{Skyrmions} 

The Goldstones are not the only excitations of the ordered phase in
the nonlinear sigma model. Since the 1970's, it has been known that
exact or approximate time independent solutions of the classical
equations of motion, when stable against quantum fluctuations, are
quantum particle excitations of the system\cite{solitons}. The
nonlinear sigma model possesses time independent solutions of a
topological nature\cite{gross1,polya1}.  These excitations are
disordered at finite length scales but relax into the N\'eel state far
away: 
    \beq 
      \lim_{|\vec x| \rightarrow \infty} \vec n=(0,0,-1) \;, \qquad
      \lim_{|\vec x| \rightarrow \infty} w=\infty \;.
    \eneq
They consist in the order parameter rotating a number of times as one
moves from infinity toward a fixed but arbitrary position in the
plane. Since two dimensional space can be thought of as an infinite 2
dimensional sphere, the excitations fall in homotopy classes of a 2D
sphere into a 2D sphere: $S^2 \rightarrow S^2$. The topological
excitations are thus defined by the number of times they map the 2D
sphere into itself. They are thus characterized by the Jacobian
    \beq
      q = \frac{1}{8 \pi} \int d^2 x \epsilon^{i j} \vec n \cdot
      \partial_i \vec n \times \partial_j \vec n \; .
    \eneq
or 
    \begin{align}
      \begin{aligned}
	q &= \frac{i}{2\pi} \int d^2 x \frac{\epsilon^{i j}\partial_i
	w \partial_j w^*}{ (1 + |w|^2)^2} \\
	&= \frac{1}{\pi} \int d^2 x \frac{\partial_z w \partial_{z^*}
	w^* - \partial_{z^*} w \partial_z w^*}{ (1 + |w|^2)^2} \;.
      \end{aligned}
    \end{align}
The number $q$ will be an integer measuring how many times the
$n$-sphere gets mapped into the infinite 2D sphere corresponding to
the plane where the spins live. If we define the space-time current
    \beq 
      J^\mu= \frac{1}{8 \pi} \epsilon^{\mu \nu \sigma } \vec n \cdot
      \partial_\nu \vec n \times \partial_\sigma \vec n =
      \frac{i}{2\pi} \epsilon^{\mu \nu \sigma} \frac{\partial_\nu w
      \partial_\sigma w^*}{(1 + |w|^2)^2} \;,
    \eneq
it is easily seen that it is conserved $\partial_\mu J^\mu = 0$ and
that the charge associated with it is our topological charge:
    \beq
      q= \int d^2x J^0 \; .
    \eneq
Thus $q$ is a conserved quantum number. These topological field
configurations were originally discovered by Skyrme\cite{skyrme} and
are called skyrmions. The conserved charge is thus the skyrmion
number.

From the expressions for the charge $q$ and for the Hamiltonian, it is
easily seen\cite{gross1,polya1} that $E \ge 4 \pi \Lambda |q| / g$. We
see that we can construct skyrmions with $q > 0$ by imposing the
condition
    \beq 
      \partial_{z^*} w =0 \;, 
    \eneq
that is $w$ is a function of $z$ only. Since the magnetization, $\vec
n$ or $w$, is a continuous function of $z$, the worst singularities it
can have are poles. The skyrmions will have a location given by the
positions of the poles or of the zeros of $w$. Far away from its
position, the field configuration will relax back to the original
N\'eel order. Therefore we have the boundary condition
$w(\infty)=\infty$, which implies
    \beq
      w=\frac{1}{\lambda^q}\prod_{i=1}^q(z-a_i) 
    \eneq
This can easily be check to have charge $q$ and energy $4 \pi \Lambda
q / g$. $\lambda^q$ is the arbitrary size and phase of the
configuration and $a_i$ are the positions of the skyrmions that
constitute the multiskyrmion configuration. The energy is independent
of the size and phase due to the conformal invariance of the
configuration. We remark that since the multiskyrmions energy is the
sum of individual skyrmion energies, the skyrmions do not interact
among themselves\cite{gross1}. An example of the explicit calculation
of the charge and energy for a diskyrmion is shown in subsection
\ref{disky}. Similarly, the multiantiskyrmion configuration can be
shown to be
    \beq
      w=\frac{1}{(\lambda^*)^q}\prod_{i=1}^q(z^*-a_i^*)
    \eneq
with charge $-q$ and energy $4 \pi \Lambda q / g$.

We have just studied the skyrmion and antiskyrmion configurations
which relax to a N\'eel ordered configuration in the $-3$ direction
far away from their positions. We shall call them $-3$-skyrmions. The
skyrmion direction is given by the boundary conditions as
$z\rightarrow\infty$. For example, $(z-a)/\lambda$ gives $n^a(\infty)
= -\delta^{3a}$, so it is a $-3$-skyrmion. The $+3$-skyrmion is
$\lambda/(z-a)$. The $+1$-skyrmion is $(z-a)/(z-b)$. The $-1$-skyrmion
is $-(z-a)/(z-b)$. The $+2$-skyrmion is $i(z-a)/(z-b)$. The
$-2$-skyrmion is $-i(z-a)/(z-b)$. Because of the rotational invariance
of the underlying theory, they are all kinematically equivalent. They
are not dynamically equivalent since a N\'eel ordered ground state has
skyrmions and antiskyrmions corresponding to its ordering direction as
excitations.

That the skyrmion configurations behave like particles follows easily
by making them time dependent and examining their dynamics. We do so
for a single skyrmion here: 
    \beq 
      w=\frac{z-a}{\lambda} \; .  
    \eneq
We make the skyrmion time dependent by allowing it to move (making its
position, $a(t)$, time dependent), and become fatter or slimmer with
time (making its size, $\lambda(t)$, time dependent). We substitute
this time dependent configuration in the Lagrangian and obtain in
subsection \ref{s-ke}:
    \beq 
      L = \frac{2 \pi \Lambda}{g} |\dot{a}|^2 - \frac{4 \pi
      \Lambda}{g} \;.
    \eneq
Since the skyrmion Lagrangian acquired a term proportional to the
skyrmion velocity squared, a kinetic energy term, we see that the
skyrmion behaves like a free particle of mass $4 \pi \Lambda / g$ with
an excitation gap of $4 \pi \Lambda / g$. The skyrmion position is a
dynamical variable. On the other hand, the conformal parameter
$\lambda$ does not have dynamics as it has infinite mass in the
thermodynamic limit, see subsection \ref{s-ke}. The conformal
parameter is thus an arbitrary constant making the skyrmion
configuration conformally invariant even when we allow time dependence
of the configuration. Even though the sigma model does not have a
microscopic length, real antiferromagnets will have a microscopic
length as a consequence of amplitude fluctuations. We thus physically
expect $|\lambda|$ to be cutoff at small values by a coherence length
$\xi$. The long distance physics is, of course, insensitive to this
cutoff.

\subsection{Skyrmion-Antiskyrmion States} \label{s-as-L}

Since the charge or skyrmion number is conserved, a configuration with
nonzero skyrmion number cannot be excited out of the ground state in
the absence of an external probe that couples to skyrmion
number. Therefore, skyrmions and antiskyrmions will be created in
equal numbers. We thus have to study the interaction between skyrmions
and antiskyrmions. A skyrmion-antiskyrmion configuration, which is not
a solution to the equations of motion, is given by
    \beq
      w=\frac{1}{\lambda^2}(z-a)(z^*-b^*) \; .
    \eneq
The energy of this static configuration is given {\it exactly} by
    \begin{align}
      \begin{aligned}
	&E_s = \frac{4 \pi \Lambda}{g} \\
	&+ \frac{\Lambda}{g} \frac{|a-b|^4}{|\lambda|^4} \int_0^\infty
	\, \frac{r K \left( \frac{4r}{(r+1)^2} \right)}{(1 + \frac{r^2
	|a-b|^4}{16 |\lambda|^4})^2 (r+1)} \, dr \; .
      \end{aligned}
    \end{align}
where $K(x)$ is the Jacobian elliptic function\cite{abste}.  We
reproduce the details of the calculation in subsection \ref{s-as}
because it has been calculated or approximated incorrectly in previous
works\cite{gross1,fat, polyab}. The skyrmion-antiskyrmion interaction,
or potential energy, is given by the difference between the static
energy $E_s$ and the sum of the energies of the isolated skyrmion and
skyrmion, $V = E_s - 8 \pi \Lambda / g$.

The skyrmion-antiskyrmion interaction has many interesting
features. As the distance between the skyrmion and antiskyrmion
becomes small compared to the size of the configuration,
$|a-b|/|\lambda| \ll 1$, their interaction is very soft; the energy
goes like
   \beq
     V \simeq - \frac{4 \pi \Lambda}{g} + \frac{\pi^2 \Lambda}{2 g}
     \frac{|a-b|^2}{|\lambda|^2} \;.
   \eneq
At short distances the skyrmions and antiskyrmions are bound by a
harmonic potential. The minimum of this classical energy occurs when
the skyrmion-antiskyrmion form a bound state with zero ``separation''
between the skyrmion and antiskyrmion, or equivalently infinite
conformal size, i.e. $|a-b|/|\lambda| = 0$. This bound state resonance
has energy $- 4 \pi \Lambda / g$, or a binding energy of $4 \pi
\Lambda / g$. Therefore the skyrmion and antiskyrmion gaps get
halved. When this skyrmion-antiskyrmion configuration has a large but
finite size, i.e. $|\lambda|/|a-b| \gg 1$, the potential between the
skyrmion and antiskyrmion is very soft and vanishes when the
configuration has arbitrarily large size. In this limit the skyrmion
and antiskyrmion do not interact despite being ``bound''.

At large distances or small size, $|a-b|/|\lambda| \gg 1$, the
interaction is approximately
    \beq
      V \simeq \frac{64 \pi \Lambda}{g} \frac{|\lambda|^4}{|a-b|^4}
      \ln \left(\frac{|a-b|}{2|\lambda|}\right) \;.
    \eneq
At large enough distances the skyrmion and antiskyrmion are almost
free and repel each other with an interaction that vanishes at
infinitely large separations. We see that the skyrmion-antiskyrmion
potential is attractive at short distances or large sizes, while at
larger distances or small size it goes to a maximum energy which is
higher than 0 and then vanishes at infinity. In order to unbind them
one has to at least supply an energy $4 \pi \Lambda / g$. Classically
one would have to supply enough energy to get over the potential
energy hump, but quantum mechanically one can, of course, tunnel
through the barrier.

Contrary to pure skyrmion or pure antiskyrmion configurations, the
skyrmion-antiskyrmion configurations are not stationary solutions of
the equations of motion. Therefore the dynamics will not be restricted
to center of mass motion alone. In order to study the dynamics of the
skyrmion and antiskyrmion configurations we allow motion of the
positions of the skyrmion, $a(t)$, and the antiskyrmion, $b(t)$, and
permit time dependence of the conformal parameter, $\lambda(t)$. We
substitute this time dependent configuration in the sigma model
Lagrangian in subsection \ref{s-as-ke} and obtain the kinetic energy
part of the Lagrangian to be
    \beq
      T = \frac{m_{ab}}{2}\left( |\dot{a}|^2 + |\dot{b}|^2\right) +
      \frac{m_{\lambda}}{2} |\dot{\lambda}|^2
    \eneq
with
    \begin{align}
      \begin{aligned}
	&m_{\lambda} = \frac{\Lambda}{2 g} \frac{|a-b|^6}{|\lambda|^6}
	\\
	&\times \int_0^\infty \frac{R^3 K \left( \frac{4 R}{(R+1)^2}
	\right) dR }{(1 + |a - b|^4 R^2 / 16 |\lambda|^4 )^2 (R + 1)}
	\\ 
	&m_{ab} = \frac{4 \Lambda}{g} \int_0^\infty \int_0^{2\pi}
	r^3 dr d\theta \times \\
	&\frac{1}{\left[1 + r^2 (r^2 + |a-b|^2 / |\lambda|^2 -
	2 r |a-b| \cos \theta) / |\lambda| \right]^2}
      \end{aligned}
    \end{align}
with $K(x)$ the Jacobian elliptic function\cite{abste}. At short
distances, or large sizes, $|a-b| \ll |\lambda|$, the $a$, $b$ and
$\lambda$ masses have the asymptotic behavior
    \beq
      m_{\lambda} \simeq \frac{4 \pi^2 \Lambda}{g} \, , \quad m_{ab}
      \simeq \frac{2 \pi \Lambda}{g} \;.
    \eneq
In this limit the mass of the skyrmion and antiskyrmion is equal to
$1/2$ the mass of an isolated skyrmion or antiskyrmion. At large
distances, or small sizes, $|a-b| \gg |\lambda|$, the masses go like
    \beq
      m_{\lambda} \simeq \frac{64 \pi \Lambda}{g}
      \frac{|\lambda|^2}{|a-b|^2} \ln \left( \frac{|a-b|^2}{4
      |\lambda|^2} \right ) \;, m_{ab} \simeq \frac{2^9 \pi
      \Lambda}{g} \frac{|\lambda|^8}{|a-b|^8} \;.
    \eneq
The Lagrangian that describes the dynamics of the
skyrmion-antiskyrmion configuration is thus
    \begin{align}
      \begin{aligned}
	L &= \frac{m_{ab}}{4} \left | \frac{d}{dt} ( a + b ) \right
	|^2 + \frac{m_{ab}}{4} \left | \frac{d}{dt} ( a - b ) \right
	|^2 \\
	&+ \frac{m_{\lambda}}{2} |\dot {\lambda}|^2 - V \left (
	\frac{|a-b|}{|\lambda|} \right ) \;.
      \end{aligned}
    \end{align}
We see that the center of mass coordinate decouples as required by the
translational invariance of the system. Contrary to the pure skyrmion
configurations, here the conformal parameter, $\lambda$, has dynamics
and is not an arbitrary parameter. That is, the skyrmion-antiskyrmion
is not conformally invariant.

\subsection{The Diskyrmion} 
\label{disky}

In the present subsection we explicitly check that a diskyrmion
configuration has charge 2 and energy equal to the sum of the energies
of the two single skyrmions that constitute the diskyrmion. This is
important because there are false claims\cite{fat} in the literature
that multiskyrmion configurations interact through logarithmic
potentials. Above we concluded that since a multiskyrmion
configuration has an energy that is the sum of the skyrmions that
constitute it, skyrmions do not interact. This was originally
concluded by Gross\cite{gross1}. In the present section we show this
by explicit calculation for the diskyrmion energy.

The diskyrmion configuration is 
    \beq
      w=\frac{1}{\lambda^2}(z-a)(z-b) \; .
    \eneq
The charge is given by
    \beq
      q=\frac{4}{\pi} \, \int \, d^2x \frac{| z - ( \tilde{a} +
      \tilde{b} ) / 2|^2}{( 1 + |z - \tilde{a}|^2 |z - \tilde{b}|^2
      )^2} \;.
    \eneq
where we have rescaled $z$ and defined $\tilde{a}=a/\lambda$ and
$\tilde{b}=b/\lambda$ in order to absorb the arbitrary size
$\lambda$. The energy is given by
    \beq
      E = \frac{16 \Lambda}{g} \, \int \, d^2x \frac{|z - ( \tilde{a}
      + \tilde{b} ) / 2|^2}{( 1 + |z - \tilde{a}|^2 |z - \tilde{b}|^2
      )^2} \;.
    \eneq
In order to calculate the energy and charge we define
    \beq
      A \equiv \frac{\tilde{a} + \tilde{b}}{2} \;, \qquad \qquad B
      \equiv \frac{\tilde{a} - \tilde{b}}{2} \;,
    \eneq
and make the change of origin $z \rightarrow z + A$, to obtain that
the energy and charge are $E=16 \Lambda I / g$ and $q = 4 I /
\pi$, where $I$ is the integral
    \begin{align}
      \begin{aligned}
        I &= \frac{1}{2} \int \frac{|z|^2 \, d z d z^*}{\left[1 +
        |z+B|^2 |z-B|^2 \right]^2} \\
	&= \frac{1}{2} \int \frac{z d z \; z^* d z^*}{\left\{1 + [z^2
        - B^2] [(z^*)^2 - (B^*)^2] \right\}^2} \\
	&= \frac{1}{8} \int \frac{d [z^2 -B^2] \; d[(z^*)^2-
	    (B^*)^2]}{\left\{1 + [z^2 - B^2] [(z^*)^2 - (B^*)^2]
	  \right\}^2} \\
	&= \frac{1}{4} \int \frac{d u \; d u^*}{\left[ 1 + |u|^2
	    \right]^2} \;.
      \end{aligned}
    \end{align}
To obtain the last equality we made the variable change $u=z^2 -B^2$.
The factor of $2$ arises because $u$ is linearly related to $z^2$, so
one must cover the $u$ complex plane twice in order to cover the $z$
complex plane once.  Going over to polar coordinates of the $u$ plane
we get
    \beq
      I=\frac{1}{2} \int \frac{r dr d \theta}{(1 + r^2)^2} =
      \frac{\pi}{2} \;.
    \eneq
We finally obtain
    \beq
      q = \frac{4}{\pi}I = 2 \;, \quad E = \frac{16
      \Lambda}{g} I = \frac{8 \pi \Lambda}{g} = \frac{4 \pi
      \Lambda}{g} q
    \eneq
as expected.

\subsection{Skyrmion Kinetic Energy} 
\label{s-ke}

When we substitute the time dependent skyrmion
    \beq
      w = \frac{z - a(t)}{\lambda(t)} \; .
    \eneq
into the sigma model Lagrangian
    \begin{align}
      \begin{aligned}	
	L &= \frac{2 \Lambda}{g} \int d^2x \frac{1}{(1 + |w|^2)^2} \\
	&\times \left( \partial_0 w \partial_0 w^* - 2 \partial_z w
	\partial_{z^*} w^* - 2 \partial_{z^*} w \partial_z w^* \right)
      \end{aligned}
    \end{align}
we obtain 
    \begin{align}
      \begin{aligned}
	L &= - \frac{4 \pi \Lambda}{g} + \frac{2 \Lambda}{g}\int d^2x
	\frac{1}{(1 + |w|^2)^2} \\
	&\times \left( \frac{| \dot{\lambda} |^2} {| \lambda |^2}
	|w|^2 + \frac{| \dot{a} |^2}{| \lambda |^2} + \frac{\dot{
	\lambda } \dot{a}^*}{| \lambda |^2} w + \frac{\dot{ \lambda
	}^* \dot{a}}{| \lambda |^2} w^* \right ) \;.
      \end{aligned}
    \end{align}
First we evaluate the integral
    \beq
      \int d^2x \frac{w}{(1 + |w|^2)^2} = |\lambda|^2 \int \frac{r^2
      e^{i \theta} dr d\theta} {(1 + r^2)^2} = 0
    \eneq
where we made the variable change $z \rightarrow z + a$, the conformal
transformation $z \rightarrow \lambda z$, and went to polar
coordinates of the complex $z$ plane. Similarly we have
    \beq
      \int d^2x \frac{w^*}{(1 + |w|^2)^2} = |\lambda|^2 \int \frac{r^2
      e^{-i \theta} dr d\theta}{(1 + r^2)^2} = 0 \;.
    \eneq
The skyrmion Lagrangian is then
    \begin{align}
      \begin{aligned}
	&L = - \frac{4 \pi \Lambda}{g} + \frac{2 \Lambda}{g} \int d^2x
	\\
	&\times \left\{ \frac{|\dot{a}|^2 }{|\lambda|^2} \frac{1}{(1 +
	|w|^2)^2} + \frac{|\dot{\lambda}|^2 }{|\lambda|^2}
	\frac{|w|^2}{(1 + |w|^2)^2} \right\} \\
	&= \frac{m_a}{2} |\dot{a}|^2 + \frac{m_\lambda}{2}
	|\dot{\lambda}|^2 -\frac{4 \pi \Lambda}{g} \;.
      \end{aligned}
    \end{align}
We see that there is a kinetic energy term for $a$ with mass
    \begin{align}
      \begin{aligned}
	m_a &= \frac{4 \Lambda}{g|\lambda|^2} \int d^2x \frac{1}{(1 +
	|w|^2)^2} \\
	&= \frac{4 \Lambda}{g} \int \frac{r dr d\theta}{(1 + r^2)^2} =
	\frac{4 \pi \Lambda}{g}
      \end{aligned}
    \end{align}
and a kinetic energy term for $\lambda$ with mass
    \begin{align}
      \begin{aligned}
        m_\lambda &= \frac{4 \Lambda}{g |\lambda|^2}\int d^2x
	\frac{|w|^2}{(1 + |w|^2)^2} \\
	&= \frac{4 \Lambda}{g}\int\frac{r^3 dr d\theta}{(1 + r^2)^2}
	\\
	&= \frac{4 \Lambda}{g} \left [ \pi \ln(1 +r^2) \Big |_0^\infty
	- \pi \right ] = \infty \;.
      \end{aligned}
    \end{align}
Since the mass for $\lambda$ is infinite, $\dot{\lambda}=0$ exactly in
order not to pay an infinite kinetic energy cost. Thus $\lambda$ is
not a dynamical variable, but a constant arbitrary parameter. This is
true classically and quantum mechanically. Quantum mechanically, the
term
    \beq
      \frac{|p_\lambda|^2}{2m_\lambda} \rightarrow 0 \qquad \text{as}
      \qquad m_\lambda \rightarrow \infty
    \eneq
with $p_\lambda$ the momentum conjugate to $\lambda$. Since the
$p_\lambda$ is the only quantity that does not commute with $\lambda$
in the Hamiltonian $H$, it immediately follows that
    \beq
      \dot{\lambda} = -\frac{i}{\hbar}\left[ \lambda, H \right] =0
    \eneq
in the infinite mass limit.

\subsection{Skyrmion-Antiskyrmion Static Energy} 
\label{s-as}

We now calculate the skyrmion number $q$ and energy $E$ of the
skyrmion-antiskyrmion static configuration
    \beq 
      w=\frac{1}{\lambda^2}(z-a)(z^*-b^*) \;.
    \eneq
The charge is given by
    \beq
      q = \frac{1}{\pi} \, \int \, d^2x \frac{|z - \tilde{b}|^2 - |z -
      \tilde{a}|^2}{(1 + |z - \tilde{a}|^2 |z - \tilde{b}|^2)^2}
    \eneq
where we have made the conformal transformation $z \rightarrow \lambda
z$ and defined $\tilde{a} = a / \lambda$ and $\tilde{b} = b / \lambda$
in order to absorb the arbitrary conformal parameter $\lambda$. The
energy is given by
    \beq
      E = \frac{4 \Lambda}{g} \, \int \, d^2x \frac{|z - \tilde{b}|^2
      + |z - \tilde{a}|^2}{(1 + |z - \tilde{a}|^2 |z -
      \tilde{b}|^2)^2} \;.
    \eneq

As in subsection \ref{disky}, in order to calculate the energy and
charge we define
    \beq
      A \equiv \frac{\tilde{a}+\tilde{b}}{2} \;, \qquad \qquad B
      \equiv \frac{\tilde{a}-\tilde{b}}{2}
    \eneq
and make the change of origin $z \rightarrow z + A$, to obtain that
the energy and charge are
    \begin{align}
      \begin{aligned}
	E &= \frac{8 \Lambda}{g} \int \frac{1}{2} \frac{|z|^2 d z d
	z^*}{\left( 1 + |z + B|^2 |z - B|^2 \right)^2} \\
	&+ \frac{4 \Lambda}{g} |B|^2 \int\frac{ d z d z^*}{\left(1 +
	|z + B|^2 |z - B|^2 \right)^2} \\
	q &= \frac{1}{2 \pi} \int d z d z^* \frac{\left( |z + B|^2 -
	|z - B|^2 \right)}{\left( 1 + |z + B|^2 |z - B|^2 \right)^2}
	\;.
      \end{aligned}
    \end{align}
The charge is easily seen to be zero. It is the subtraction of two
integrals. If on the second integral we take $z \rightarrow -z$, it
becomes equal to the first and thus cancels it upon
subtraction. Therefore $q = 0$ for the skyrmion-antiskyrmion
configuration as expected. The first term in the energy was evaluated
in subsection \ref{disky}. We thus have
    \beq
      E = \frac{4 \pi \Lambda}{g} + \frac{4 \Lambda}{g} |B|^2 \int
      \frac{ d z d z^*}{\left[1 + |z + B|^2 |z - B|^2 \right]^2} \;.
    \eneq
Making the variable change $u=z^2 - B^2$, we get
    \begin{align}
      \begin{aligned}
        &E = \frac{4 \pi \Lambda}{g} + \frac{2 \Lambda}{g} |B|^2 \int
	\frac{ d u d u^*}{|u + B^2|\left[1 + |u|^2\right]^2} \\
        &= \frac{4 \pi \Lambda}{g} + \frac{4 \Lambda}{g} |B|^2 \int r
        dr d\theta \\
	&\times \frac{1}{(1 + r^2)^2 \sqrt{r^2 + |B|^4 + 2 r |B|^2
	\cos \theta}}
      \end{aligned}
    \end{align}
where for the last equality we went to polar coordinates of the
complex $u$ plane. Now
    \begin{align} 
      &\int_0^{2\pi} \frac{ d\theta}{\sqrt{r^2 + |B|^4 + 2 r |B|^2
      \cos \theta}} \\
      \nonumber
      &= 2 \int_0^{\pi}\frac{ d\theta}{\sqrt{r^2 + |B|^4 + 2 r |B|^2
      \cos \theta}} \\
      \nonumber
      &= 2 \int_0^{\pi}\frac{ d\theta}{\sqrt{r^2 + |B|^4 + 2r|B|^2 - 4
      r |B|^2 \sin^2(\theta/2)}} \\
      \nonumber
      &= 4 \int_0^{\pi/2}\frac{ d\phi}{\sqrt{r^2 + |B|^4 + 2r|B|^2 - 4
      r |B|^2 \sin^2 \phi}} \\
      \nonumber
      &= \frac{4}{r + |B|^2}K\left( \frac{4 r |B|^2}{( r + |B|^2 )^2}
      \right)
    \end{align}
where $K(x)$ is the Jacobian elliptic function\cite{abste}. After
making the variable change $R = |B|^2 r$, we then have
    \begin{align} 
      \begin{aligned}
        E &= \frac{4 \pi \Lambda}{g} + \frac{16 \Lambda}{g} |B|^4
	\int_0^\infty \frac{ K \left( \frac{ 4 R }{ (R + 1)^2 }
	\right) R dR }{ (R + 1)(1 + |B|^4 R^2)^2 } \\
	&= \frac{4 \pi \Lambda}{g} + \frac{\Lambda}{g} \frac{|a -
	b|^4}{|\lambda|^4} \int_0^\infty K \left( \frac{ 4 R }{ (R +
	1)^2 } \right) R dR \\
	&\times \frac{1}{ [R + 1][1 + |a-b|^4 R^2/(16 |\lambda|^4)]^2
	}\;.
      \end{aligned}
    \end{align}
Asymptotic approximations yield to leading order
    \begin{align}
      \begin{aligned}
        E &\simeq \frac{4 \pi \Lambda}{g} + \frac{\pi^2 \Lambda}{2 g}
        \frac{|a - b|^2}{|\lambda|^2} &\; \text{ for } \; \frac{|a -
        b|}{|\lambda|} \ll 1& \\
	E &\simeq \frac{8 \pi \Lambda}{g} + \frac{64 \pi \Lambda}{g}
	\frac{|\lambda|^4}{|a-b|^4} \\
	&\times \ln \left(\frac{|a - b|}{2|\lambda|} \right) &\;
	\text{ for } \; \frac{|a - b|}{|\lambda|} \gg 1& \;.
      \end{aligned}
    \end{align}

\subsection{Skyrmion-Antiskyrmion Kinetic Energy} 
\label{s-as-ke}

    \begin{widetext} 
We now move to determine the kinetic energy of the time dependent 
skyrmion-antiskyrmion configuration:
      \beq
        w = \frac{[z-a(t)][z^* - b^*(t)]}{\lambda^2(t)} \; .
      \eneq
we substitute
      \begin{align} 
	\begin{aligned}
          \partial_0 w \partial_0 w^* &= 4
          \frac{|\dot{\lambda}|^2}{|\lambda|^2} |w|^2 +
          \frac{|\dot{a}|^2}{|\lambda|^4} |z - b|^2 +
          \frac{|\dot{b}|^2}{|\lambda|^4} |z - a|^2 + 2
          \frac{\dot{\lambda} \dot{a}^*}{|\lambda|^2 \lambda^*} (z -
          b) \, w + 2 \frac{\dot{\lambda}^* \dot{a}}{|\lambda|^2
          \lambda} (z^* - b^*) \, w^* \\
	  &+ 2 \frac{\dot{\lambda} \dot{b}}{|\lambda|^2 \lambda^*}
          (z^* - a^*) \, w + 2 \frac{\dot{\lambda}^*
          \dot{b}^*}{|\lambda|^2 \lambda} (z - a) \, w^* +
          \frac{\dot{a}\dot{b}}{|\lambda|^4} (z^* - a^*)(z^* - b^*) +
          \frac{\dot{a}^*\dot{b}^*}{|\lambda|^4} (z - a)(z - b)
	\end{aligned}
      \end{align}
into the kinetic term of the sigma model Lagrangian
      \beq
        L = \frac{2 \Lambda}{g}\int d^2x \frac{\partial_0 w \partial_0
	w^* - 2 \, \partial_z w \partial_{z^*} w^* - 2 \,
	\partial_{z^*} w \partial_z w^*}{(1 + |w|^2)^2}
      \eneq
to obtain 
      \begin{align}
	\begin{aligned}
	  \label{lag}
	  L &= \int d^2x \left( \frac{8 \Lambda}{g}
	  \frac{|\dot{\lambda}|^2}{|\lambda|^2} \frac{|w|^2}{(1 +
	  |w|^2)^2} + \frac{2 \Lambda}{g}
	  \frac{|\dot{a}|^2}{|\lambda|^4} \frac{|z-b|^2}{(1 +
	  |w|^2)^2} + \frac{2 \Lambda}{g}
	  \frac{|\dot{b}|^2}{|\lambda|^4} \frac{|z - a|^2}{(1 +
	  |w|^2)^2} \right. \\
	  &+ \frac{4 \Lambda}{g} \frac{\dot{\lambda}
	  \dot{a}^*}{|\lambda|^2 \lambda^*} \frac{(z - b) w}{(1 +
	  |w|^2)^2} + \frac{4 \Lambda}{g} \frac{\dot{\lambda}^*
	  \dot{a}}{|\lambda|^2 \lambda} \frac{(z^* - b^*) \, w^*}{(1 +
	  |w|^2)^2} + \frac{4 \Lambda}{g} \frac{\dot{\lambda}
	  \dot{b}}{|\lambda|^2 \lambda^*} \frac{(z^* - a^*) \, w}{(1 +
	  |w|^2)^2} \\
	  &+ \left. \frac{4 \Lambda}{g} \frac{\dot{\lambda}^*
	  \dot{b}^*}{|\lambda|^2 \lambda} \frac{(z - a) \, w^*}{(1 +
	  |w|^2)^2} + \frac{2 \Lambda}{g} \frac{\dot{a}
	  \dot{b}}{|\lambda|^4} \frac{(z^* - a^*)(z^* - b^*)}{(1 +
	  |w|^2)^2} + \frac{2 \Lambda}{g} \frac{\dot{a}^*
	  \dot{b}^*}{|\lambda|^4} \frac{(z - a)(z - b)}{(1 + |w|^2)^2}
	  \right) \;.
	\end{aligned}
      \end{align}
    \end{widetext}

We now evaluate term by term of the Lagrangian. The first is
    \begin{align}
      \begin{aligned}
	L_1 &= \frac{8|\dot{\lambda}|^2}{g|\lambda|^2} \int d^2x
	\frac{|w|^2}{(1 + |w|^2)^2} \\
	&= \frac{4|\dot{\lambda}|^2}{g}\int dz dz^*
	\frac{|z-\tilde{a}|^2|z-\tilde{b}|^2}{(1 +
	|z-\tilde{a}|^2|z-\tilde{b}|^2)^2}
      \end{aligned}
    \end{align}
where we made the conformal transformation $z \rightarrow \lambda z$,
and defined $\tilde{a}= a/\lambda$ and $\tilde{b}= b/\lambda$. As in
subsection \ref{disky}, we define
    \beq
      A \equiv \frac{\tilde{a}+\tilde{b}}{2} \;, \qquad \qquad \quad B
      \equiv \frac{\tilde{a}-\tilde{b}}{2}
    \eneq
and make the change of origin $z \rightarrow z + A$, to obtain
    \begin{align}
      \nonumber
      &L_1 = \frac{4 \Lambda}{g} |\dot{\lambda}|^2 \int dz dz^*
      \frac{|z - B|^2|z + B|^2}{(1 + |z - B|^2|z + B|^2)^2} \\
      \nonumber
      &= \frac{2 \Lambda}{g} |\dot{\lambda}|^2 \int du du^*
      \frac{|u|^2}{(1 + |u|^2)^2|u + B^2|} \\
      \nonumber
      &= \frac{4 \Lambda}{g} |\dot{\lambda}|^2 \int \frac{r^3 dr
      d\theta}{(1 + r^2)^2 \sqrt{r^2 + |B|^4 + 2 r |B|^2 \cos \theta}}
      \\
      \nonumber
      &= \frac{16 \Lambda}{g} |\dot{\lambda}|^2 \int \frac{r^3 K
      \left(\frac{4 r |B|^2}{(r + |B|^2)^2}\right) dr }{(1 + r^2)^2 (r
      + |B|^2)} \\
      \nonumber
    \end{align}
    \begin{align}
      \nonumber
      &= \frac{\Lambda}{4 g} \frac{|\dot{\lambda}|^2}{|\lambda|^6} |a
      - b|^6 \int_0^\infty \frac{R^3 K \left( \frac{4 R}{(R + 1)^2}
      \right) dR }{(1 + |a - b|^4 R^2 / 16 |\lambda|^4)^2 (R + 1)} \\
      &\equiv m_{\lambda}\frac{|\dot{\lambda}|^2}{2}
    \end{align}
where the second line follows from the variable change $u = z^2 -
B^2$, and the third by going to polar coordinates in the complex $u$
plane. $K(x)$ is the Jacobian elliptic function\cite{abste}.  We have
    \begin{align}
      \begin{aligned}
        L_1 &\simeq \frac{2 \pi^2 \Lambda}{g} |\dot{\lambda}|^2
	&\text{ for } \frac{|a-b|}{|\lambda|} \ll 1& \\
        L_1 &\simeq \frac{32 \pi \Lambda}{g} \frac{|\dot{\lambda}|^2
        |\lambda|^2}{|a-b|^2} \\
	&\times \ln \left( \frac{|a - b|^2}{4 |\lambda|^2} \right)
	&\text{ for } \frac{|a - b|}{|\lambda|} \gg 1& \;.
    \end{aligned}
  \end{align}

The second term of the Lagrangian (\ref{lag}) is
    \begin{align}
      \begin{aligned}
	\nonumber
        L_2 &= \frac{2|\dot{a}|^2}{g|\lambda|^4}\int d^2x
	\frac{|z-b|^2}{(1 + |w|^2)^2} \\
	&= \frac{|\dot{a}|^2}{g} \int dz dz^* \frac{|z -
	\tilde{b}|^2}{(1 + |z - \tilde{a}|^2 |z - \tilde{b}|^2)^2} \\
	&= \frac{|\dot{a}|^2}{g} \int dz dz^* \frac{|z|^2}{(1 + |z -
	2B|^2 |z|^2)^2} \\
	&= \frac{2|\dot{a}|^2}{g} \int_0^\infty \int_0^{2 \pi}
	\frac{r^3 dr d\theta}{ \left[1 + r^2(r^2 + 4|B|^2 - 4 r |B|
	\cos \theta) \right]^2}
      \end{aligned}
    \end{align}
    \beq
      \equiv \frac{|\dot{a}|^2}{2} m_a \left( \frac{|a -
      b|}{|\lambda|} \right)
    \eneq
where we made the shift $z \rightarrow z + A - B$ from the second to
the third line. The mass for the $a$ coordinate is defined through the
integral
    \begin{widetext}
      \beq
        m_a \left( \frac{|a-b|}{|\lambda|} \right) =
	\frac{4}{g}\int_0^\infty \int_0^{2\pi} \frac{r^3 dr
	d\theta}{\left[1 + r^2(r^2 + 4|B|^2 -
	4r|B|\cos\theta)\right]^2}
      \eneq
with $B = (a - b)/(2 \lambda)$. For $|a - b| \ll |\lambda|$, we have
      \beq
        m_a = \frac{2\pi}{g} \;.
      \eneq
For $|a - b| \gg |\lambda|$ with $r = |B| \, R$, we have
      \beq
        m_a \left( \frac{|a - b|}{|\lambda|} \right) = \frac{4}{g
        |B|^4} \int_0^\infty \int_0^{2 \pi} \frac{R^3 dR
        d\theta}{\left[(1 / |B|^4) + R^4 + 4 R^2 - 4 R^3 \cos\theta)
        \right]^2} \simeq \frac{ 2^9 \pi |\lambda|^8}{g |a - b|^8} \;.
      \eneq
    \end{widetext}
Similarly the third term of the Lagrangian (\ref{lag}) is
    \beq
      L_3 = \frac{|\dot{b}|^2}{2} m_b \left( \frac{|a - b|}{|\lambda|}
      \right)
    \eneq
with
    \beq
      m_b \left( \frac{|a - b|}{|\lambda|} \right) = m_a \left(
      \frac{|a - b|}{|\lambda|} \right) \;.
    \eneq
The rest of the terms of the Lagrangian (\ref{lag}) come out to be
zero by rotational invariance in the complex $z$ plane.

\end{document}